\documentclass[aps,prl,twocolumn,amsfonts,amssymb,amsmath,floats,floatfix,showpacs,preprintnumbers,superscriptaddress]{revtex4}
\pdfoutput=1
\usepackage{graphicx}
\usepackage{dcolumn}
\usepackage{bm}
\usepackage{epsfig,graphicx}
\usepackage{graphics}
\usepackage{multirow}

\usepackage{times}

\begin{document}

\title{Charge order driven by Fermi-arc instability in Bi$_2$Sr$_{2-x}$La$_{x}$CuO$_{6+\delta}$} 

\author{R. Comin}
\affiliation{Department of Physics {\rm {\&}} Astronomy, University of British Columbia, Vancouver, British Columbia V6T\,1Z1, Canada}
\author{A. Frano}
\affiliation{Max Planck Institute for Solid State Research, Heisenbergstrasse 1, D-70569 Stuttgart, Germany}
\affiliation{Helmholtz-Zentrum Berlin f\"{u}r Materialien und Energie, Albert-Einstein Stra{\ss}e 15, D-12489 Berlin, Germany}
\author{M. M. Yee}
\affiliation{Department of Physics, Harvard University, Cambridge, MA 02138, U.S.A.}
\author{Y. Yoshida}
\affiliation{National Institute of Advanced Industrial Science and Technology (AIST), Tsukuba, 305-8568, Japan}
\author{H. Eisaki}
\affiliation{National Institute of Advanced Industrial Science and Technology (AIST), Tsukuba, 305-8568, Japan}
\author{E. Schierle}
\affiliation{Helmholtz-Zentrum Berlin f\"{u}r Materialien und Energie, Albert-Einstein Stra{\ss}e 15, D-12489 Berlin, Germany}
\author{E.\,Weschke}
\affiliation{Helmholtz-Zentrum Berlin f\"{u}r Materialien und Energie, Albert-Einstein Stra{\ss}e 15, D-12489 Berlin, Germany}
\author{R. Sutarto}
\affiliation{Canadian Light Source, University of Saskatchewan, Saskatoon, Saskatchewan S7N\,2V3, Canada}
\author{F. He}
\affiliation{Canadian Light Source, University of Saskatchewan, Saskatoon, Saskatchewan S7N\,2V3, Canada}
\author{A. Soumyanarayanan}
\affiliation{Department of Physics, Harvard University, Cambridge, MA 02138, U.S.A.}
\author{Yang He}
\affiliation{Department of Physics, Harvard University, Cambridge, MA 02138, U.S.A.}
\author{M. Le Tacon}
\affiliation{Max Planck Institute for Solid State Research, Heisenbergstrasse 1, D-70569 Stuttgart, Germany}
\author{I.\,S.\,Elfimov}
\affiliation{Department of Physics {\rm {\&}} Astronomy, University of British Columbia, Vancouver, British Columbia V6T\,1Z1, Canada}
\affiliation{Quantum Matter Institute,University of British Columbia, Vancouver, British Columbia V6T\,1Z4, Canada}
\author{J. E. Hoffman}
\affiliation{Department of Physics, Harvard University, Cambridge, MA 02138, U.S.A.}
\author{G. A. Sawatzky}
\affiliation{Department of Physics {\rm {\&}} Astronomy, University of British Columbia, Vancouver, British Columbia V6T\,1Z1, Canada}
\affiliation{Quantum Matter Institute,University of British Columbia, Vancouver, British Columbia V6T\,1Z4, Canada}
\author{B. Keimer}
\affiliation{Max Planck Institute for Solid State Research, Heisenbergstrasse 1, D-70569 Stuttgart, Germany}
\author{A. Damascelli}
\email{damascelli@physics.ubc.ca}
\affiliation{Department of Physics {\rm {\&}} Astronomy, University of British Columbia, Vancouver, British Columbia V6T\,1Z1, Canada}
\affiliation{Quantum Matter Institute,University of British Columbia, Vancouver, British Columbia V6T\,1Z4, Canada}

\begin{abstract}
An understanding of the nature of superconductivity in cuprates has been hindered by the apparent diversity of intertwining electronic orders in these materials. Here we combine resonant X-ray scattering (REXS), scanning-tunneling microscopy (STM), and angle-resolved photoemission spectroscopy (ARPES) to observe a charge order that appears consistently in surface and bulk, as well as  momentum and real space, with the Bi$_2$Sr$_{2-x}$La$_{x}$CuO$_{6+\delta}$ cuprate family. The observed wavevector rules out simple antinodal nesting in the single particle limit, but matches well with a phenomenological model of a many-body instability of the Fermi arcs. Combined with earlier observations in other cuprate families, these findings suggest the existence of a generic charge-ordered state in underdoped cuprates, and uncover its connection to the pseudogap regime.
\end{abstract}

\maketitle

Since the discovery of cuprate high-temperature superconductors, several unconventional phenomena have been observed in the region of the phase diagram located between the strongly localized Mott insulator at zero doping and the itinerant Fermi-liquid state that emerges beyond optimal doping \cite{Bonn_review,FournierNP,kaminski2002,Hoffman2002,howald2003,vershinin2004,wise2008,hanaguri2004,tranquada1995,vZimmermann_1998,abbamonte2005,shen2005,Li2008,lawler2010,hashimoto2010,he2011,wu2011,ghiringhelli2012,chang2012,LeTacon2013}.
The so-called `pseudogap' opens at the temperature ${T}^{*}$ and obliterates the Fermi surface at the antinodes (AN) of the d-wave superconducting gap function, leaving behind disconnected ``Fermi arcs'' centered around the nodes. In addition, charge order has been observed on the surface of Bi- and Cl-based compounds \cite{Hoffman2002,howald2003,vershinin2004,wise2008,hanaguri2004}, in the bulk of La-based compounds \cite{tranquada1995,vZimmermann_1998,abbamonte2005}, and most recently in YBa$_{2}$Cu$_{3}$O$_{6+\delta}$ (YBCO) \cite{wu2011,ghiringhelli2012,chang2012,LeTacon2013}, indicating this might be the leading instability in underdoped cuprates. The similarity between the observed charge ordering wavevector and the antinodal nesting vector of the high-temperature Fermi surface has prompted suggestions that a conventional Peierls-like charge-density-wave (CDW) might be responsible for the opening of the pseudogap \cite{hanaguri2004,shen2005,wise2008,chang2012}. We use complementary bulk/surface techniques to examine the validity of this scenario, and explore the connection between charge ordering and fermiology.
\begin{figure}[t!]
\centerline{\epsfig{file=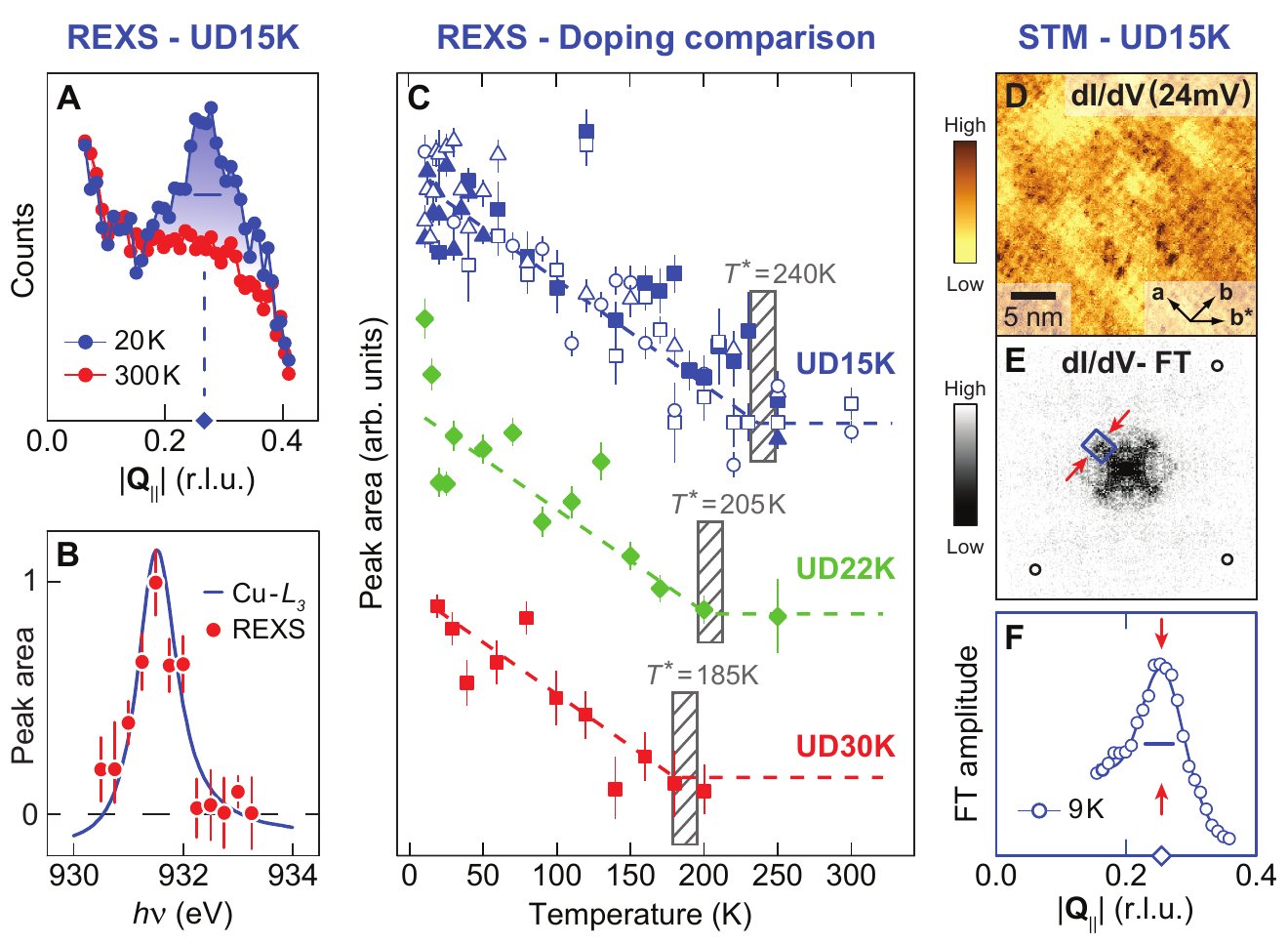,clip=,angle=0,width=1\linewidth}}
\vspace{-2mm}
\caption{REXS and STM comparison on Bi2201. (\textbf{A}) Bi2201 low- and high-temperature scans of the in-plane momentum $ {Q}_{\parallel} $, showing the emergence of a charge-order (CO) peak around $ {Q}_{\mathrm{CO}} \!\simeq\! 0.265 $. (\textbf{B}) Resonance profile at ${Q}_{\mathrm{CO}}$ after background subtraction (red markers), superimposed onto the Cu-${L}_{3}$ absorption edge (blue line). (\textbf{C}) Temperature dependence of the CO-peak area for the three Bi2201 doping levels investigated; dashed lines are linear fits, with baseline marking the zero intensity position for each dataset. Hollow (full) markers for UD15K correspond to positive (negative) wavevectors ${Q}_{\parallel}\!>\!0$ (${Q}_{\parallel}\!<\!0$), for different samples. The pseudogap temperature ${T}^{*}$ (grey boxes) is from Knight shift measurements \cite{Kawasaki2010}. (\textbf{D}) Bi2201 $\mathrm{d}I/\mathrm{d}V$ map, taken at 24\,mV bias over a 29\,nm region (9\,K, -200\,mV, and 250\,pA). A charge modulation with period $\sim\!4{a}_{0}$ is clearly seen in real space. (\textbf{E}) FT of (\textbf{D}), after four-fold symmetrization [black circles mark the position of the Bragg vectors $(\pm 1, 0)$ and $(0, \pm 1)$]. Highlighted by the blue box is the region corresponding to the linecut in (\textbf{F}), whose peak structure is suggestive of a periodic modulation with wavevector $ {Q}_{\mathrm{CO}} \!\simeq\! 0.248 $.}
\label{REXS_fig}
\end{figure}

By applying a suite of complementary tools to a single cuprate material, Bi$_2$Sr$_{2-x}$La$_{x}$CuO$_{6+\delta}$ (Bi2201), we reveal that the charge order in this system emerges just below ${T}^{*}$, and that its wavevector corresponds to the Fermi arc tips rather than the antinodal nesting vector. We quantify the Fermi surface using ARPES, and we look for charge modulations along the Cu-O bond directions in both real- and reciprocal-space, using STM and REXS. The single-layer Bi2201 is well suited to this purpose owing to: (i) its two-dimensionality and high degree of crystallinity \cite{phil,Rosen_Comin}, and (ii) the possibility of probing the temperature evolution across ${T}^{*}$, which is better-characterized \cite{hashimoto2010,he2011} and more accessible than in bilayer systems. This study, by bringing together three different techniques on the same material belonging to the Bi-family, and together with related observations on La- and Y-based compounds, also suggests the ubiquity of charge ordering in underdoped cuprates [see also the subsequent report for bilayer Bi2212 \cite{daSilvaNeto2013}].

REXS uses X-ray photons to exchange momentum with the electrons and the ionic lattice, in order to gain information on the electronic charge distribution. As opposed to conventional X-ray diffraction, which is widely used for structural studies, in REXS the photon energy is tuned to resonance with one of the element-specific absorption lines. This results in a strong enhancement of the sensitivity to the valence electrons, and allows the detection of very small variations in the electronic density profile within the CuO${}_{2}$ planes \cite{abbamonte2005}, which are difficult to determine using nonresonant methods. We performed REXS measurements along the tetragonal crystallographic \textbf{a} and \textbf{b} axes, with the corresponding reciprocal axes labeled ${\mathbf{Q}}_{H}$ and ${\mathbf{Q}}_{K}$, for 3 doping levels of Bi2201 \cite{Materials_Methods}. Due to the near-equivalence of ${\mathbf{Q}}_{H}$ and ${\mathbf{Q}}_{K}$, we will hereafter use the common notation ${\mathbf{Q}}_{\parallel}$, and define reciprocal lattice units (r.l.u.) for momentum axes as: $ 2 \pi / {a}_{0}\!=\!2 \pi / {b}_{0}\!=\!1 $, with ${a}_{0}\!\simeq\!{b}_{0}\!\simeq\!3.86 \mathrm{\AA}$. Figure \ref{REXS_fig}A shows REXS scans at high (300\,K) and low temperature (20\,K) acquired on a UD15K sample near the Cu-${L}_{3}$ absorption peak at a photon energy $h \nu \!=\!931.5$\,eV. An enhancement of scattering intensity, in the form of a broad peak, is clearly visible at 20\,K at $ \vert {\mathbf{Q}}_{\parallel} \vert \!=\!0.265\!\pm\!0.01$, whereas at 300\,K it disappears into the featureless background (dominated by fluorescence). By subtracting the latter, we can study the dependence of the low-temperature feature on photon energy, which reveals its resonant behavior at the Cu-${L}_{3}$ edge (Fig.\,\ref{REXS_fig}B). The resonant enhancement, together with the absence of features at the La-${M}_{5}$ absorption edge (fig.\,S\ref{tth_pol_edge}), demonstrates that the peak originates from charge order (CO) occurring in the CuO${}_{2}$ planes. Furthermore, the gradual dependence of the peak intensity on the out-of-plane component of the wavevector ${Q}_{\perp}$ is similar to observations in YBCO \cite{chang2012}, and indicative of short coherence along the \textbf{c} axis. Figure \ref{REXS_fig}C shows the temperature evolution of the CO peak in REXS: there is a clear onset temperature ${T}_{\mathrm{CO}}$, but we cannot conclusively determine if ${T}_{\mathrm{CO}}$ corresponds to a sharp phase boundary. Although the charge modulation breaks translational symmetry, the system lacks long-range order as evidenced by the short correlation length (${\xi}_{\mathrm{CO}}\!\sim\!20 - 30\mathrm{\AA}$). The latter evolves only weakly with doping and temperature (fig.\,S\ref{Pos_width}), and therefore suggests either strong disorder or substantial fluctuations persisting down to low temperatures \cite{Robertson2006}. In either case, the convergence of ${T}_{\mathrm{CO}}$ and ${T}^{*}$ for all doping levels suggests an intimate relationship between the CO and the PG correlations.
\begin{figure*}[t!]
\centerline{\epsfig{file=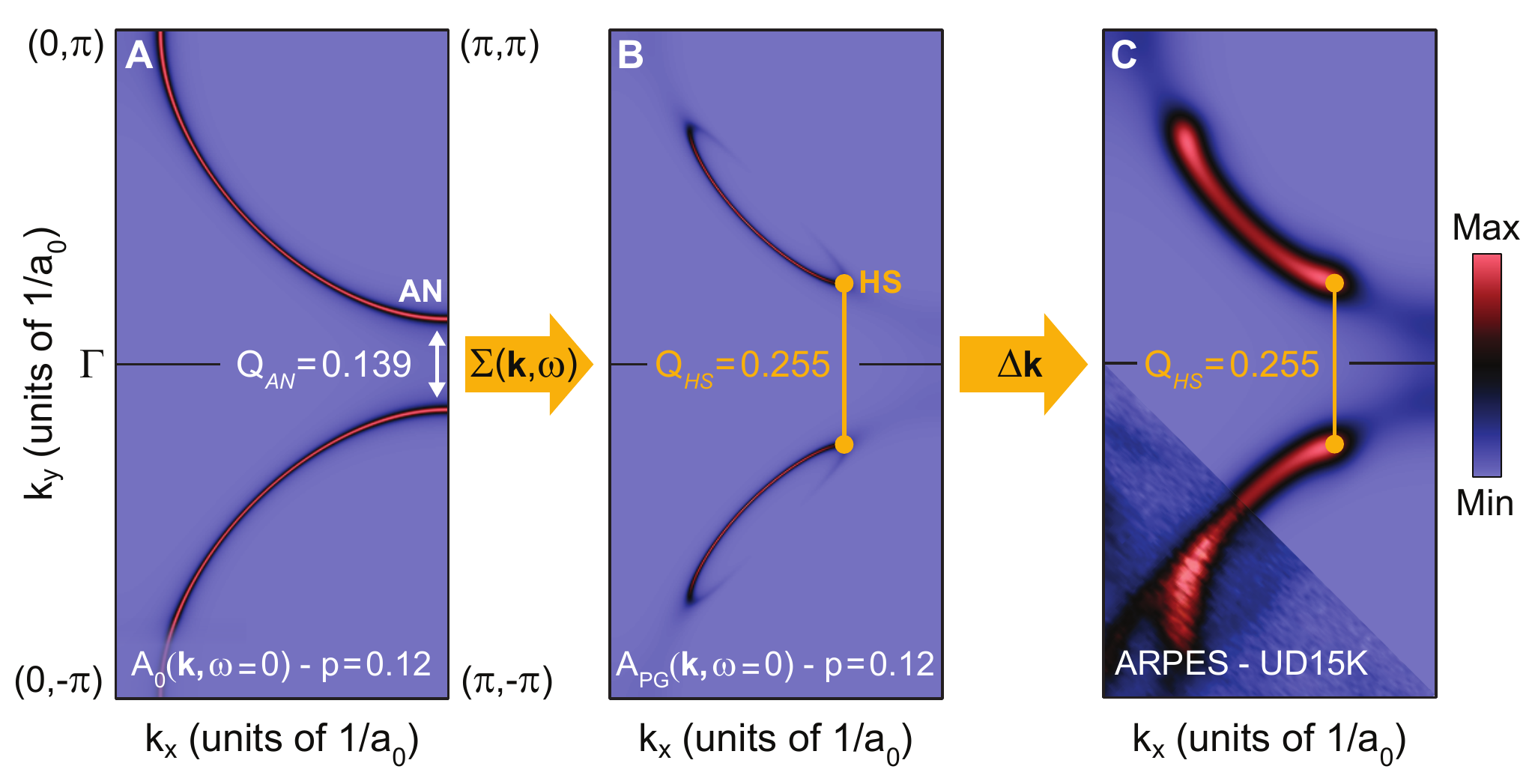,clip=,angle=0,width=0.65\linewidth}}
\vspace{-2mm}
\caption{ARPES and theory comparison on Bi2201. Modeled evolution of the Fermi surface for hole-doping $p\!=\!0.12$ from (\textbf{A}) the non-interacting to (\textbf{B}) the interacting case, via the inclusion of the self-energy ${\Sigma}_{\mathrm{PG}} (\mathbf{k}, \omega)$. A further Gaussian smearing (\textbf{C}), with $\Delta {k}_{x}\!=\!\Delta {k}_{y}\!=\!0.03 \: \pi / a$ representing the effective experimental resolution, allows comparison between the calculated and measured Fermi surface from UD15K Bi2201 \cite{phil,Rosen_Comin}. Also shown is the progression from antinodal (AN) nesting at $Q_{AN}$ -- highlighted by the white arrow -- to the $Q_{HS}$-vector associated with the tips of the Fermi arcs (hot-spots, HS) -- marked by the gold connector.}
\label{Susc_intro_fig}
\end{figure*}

STM is used to detect the charge distribution in real space, by scanning an atomically sharp tip over the cleaved Bi2201 surface and mapping the differential tunneling conductance $\mathrm{d}I/\mathrm{d}V(\mathbf{r},V)$, which is proportional to the local density of states at energy $\epsilon\!=\!eV$.
Here we apply STM to the same UD15K sample studied by REXS. The map of $\mathrm{d}I/\mathrm{d}V(\mathbf{r},V\!=\!24\,\mathrm{mV})$ in Fig.\,\ref{REXS_fig}D shows an incommensurate charge modulation along the \textbf{a} and \textbf{b} axes -- consistent with either a disordered checkerboard or stripe modulation \cite{Robertson2006}. The Fourier transform of $\mathrm{d}I/\mathrm{d}V(\mathbf{r},V)$ (Fig.\,\ref{REXS_fig}E) and associated linecut (Fig.\,\ref{REXS_fig}F) quantify the CO peak at $ \vert {\mathbf{Q}}_{\parallel} \vert \!=\!0.248\!\pm\!0.01$. This is in good agreement with ${Q}_{\mathrm{CO}}$ from REXS and also with ${Q}_{\mathrm{CO}}$ recently reported in the context of phonon anomalies in Bi2201 \cite{Bonnoit2012}. Furthermore, the feature found in STM has a correlation length ${\xi}_{\mathrm{CO}}\!\sim\!28\,\mathrm{\AA}$, again in agreement with REXS. A summary of the REXS and STM results is presented in Table\,\ref{Pos_width_table}. We therefore arrive at the empirical convergence of a charge order which onsets right below ${T}^{*}$ (REXS), and whose wavevector is consistent on surface (STM) and bulk (REXS).  
\begin{table}[b!]
\begin{center}
\begin{tabular}{c|c|c|c|c|c}
\hline
\hline
Technique & Sample & \multicolumn{4}{|c}{Parameters} \\
 & & \multicolumn{4}{c}{} \\
 & & ${Q}_{\mathrm{CO}}$ (r.l.u.) & ${\xi}_{\mathrm{CO}}$ (\AA) & ${T}_{\mathrm{CO}}$ (K) & ${T}^{*}$ (K) \\
\hline 
\multirow{2}{*}{REXS} & UD30K & $ 0.243 \pm 0.01 $ & $ 21 \pm 3 $ & $ 180 \pm 30 $ & $ 185 \pm 10 $ \\
 & UD22K & $ 0.257 \pm 0.01 $ & $ 23 \pm 3 $ & $ 202 \pm 20 $ & $ 205 \pm 10 $ \\
 & UD15K & $ 0.265 \pm 0.01 $ & $ 26 \pm 3 $ & $ 237 \pm 10 $ & $ 240 \pm 10 $ \\
\hline 
STM & UD15K & $ 0.248 \pm 0.01 $ & $ 28 \pm 2 $ & n/a & $ 240 \pm 10 $ \\
\hline 
ARPES & UD15K & $ 0.255 \pm 0.01 $ & n/a & n/a & $ 240 \pm 10 $ \\
\hline 
\hline
\end{tabular}
\end{center}
\caption{Comparative summary for the charge-order peak parameters as seen with REXS and STM for the various doping levels. For ARPES, the value listed here corresponds to the observed ${Q}_{HS}$, also shown in Fig.\,\ref{Susc_doping_fig}C. The pseudogap temperature ${T}^{*}$ (grey boxes in Fig.\,\ref{REXS_fig}C) is from Knight shift measurements \cite{Kawasaki2010}.}
\label{Pos_width_table}
\end{table}

The next step is to link the universal surface and bulk charge order to the fermiology. We quantify and clarify this connection by using ARPES to map the Fermi surface on the same UD15K Bi2201 sample studied by REXS and STM \cite{phil,Rosen_Comin}. In a similar context, the ARPES-derived `octet model' in the interpretation of quasiparticle scattering as detected by STM \cite{McElroy_2003}, is a successful example of such a connection, and demonstrates the importance of low-energy particle-hole scattering processes across the `pseudogapped' Fermi surface.

From the raw ARPES data (Fig.\,\ref{Susc_intro_fig}C \cite{phil}), we deduce that the charge ordering wavevector connects the Fermi arc tips, not the antinodal Fermi surface sections as it had been previously assumed \cite{shen2005,wise2008,chang2012}. To better understand the empirical link between charge order and fermiology, we first derive the non-interacting band structure by fitting the ARPES-measured spectral function ${A}_{\mathrm{exp}} (\mathbf{k}, \omega)$ to a tight-binding model \cite{phil,Rosen_Comin,Materials_Methods}. The corresponding Fermi surface is shown in Fig.\,\ref{Susc_intro_fig}A for the case of $p\!=\!0.12$, equivalent to UD15K \cite{Ando2000}. The AN nesting, marked by the white arrow, yields an ordering wavevector ${Q}_{AN}\!\sim\!0.139$, in disagreement with the REXS/STM average value ${Q}_{CO}\!\sim\!0.256$. To account for the suppression of antinodal zero-energy quasiparticle excitations -- a hallmark of the pseudogap (PG) fermiology -- we construct a model spectral function ${A}_{\mathrm{PG}} (\mathbf{k}, \omega)$ with an appropriate self-energy ${\Sigma}_{\mathrm{PG}} (\mathbf{k}, \omega)$, which combines the features found from exact diagonalization of the Hubbard model \cite{Eder} with the doping-dependent parameters introduced in Ref.\,\onlinecite{YRZ} (see Supplementary Note 3 for more details). Figure \ref{Susc_intro_fig}B shows how the non-interacting Fermi surface is transformed by the action of our ${\Sigma}_{\mathrm{PG}} (\mathbf{k}, \omega)$, and also highlights the concurrent shift in the smallest-$Q$ zero-energy particle-hole excitation (gold connectors). The interacting spectral function ${A}_{\mathrm{PG}} (\mathbf{k}, \omega)$ used here is tuned to optimize the match with the corresponding ARPES data \cite{phil,Rosen_Comin}; after accounting for instrumental resolution $\Delta \mathbf{k}$, the agreement with the experimental data is excellent, as shown in Fig.\,\ref{Susc_intro_fig}C. The vector connecting the tips of the Fermi arcs, called hot-spots (HS), is found to be ${Q}_{HS}\!\sim\!0.255$, closely matching the experimental values of ${Q}_{CO}$ found for the UD15K sample (see also Table\,\ref{Pos_width_table}).

\begin{figure}[b!]
\centerline{\epsfig{file=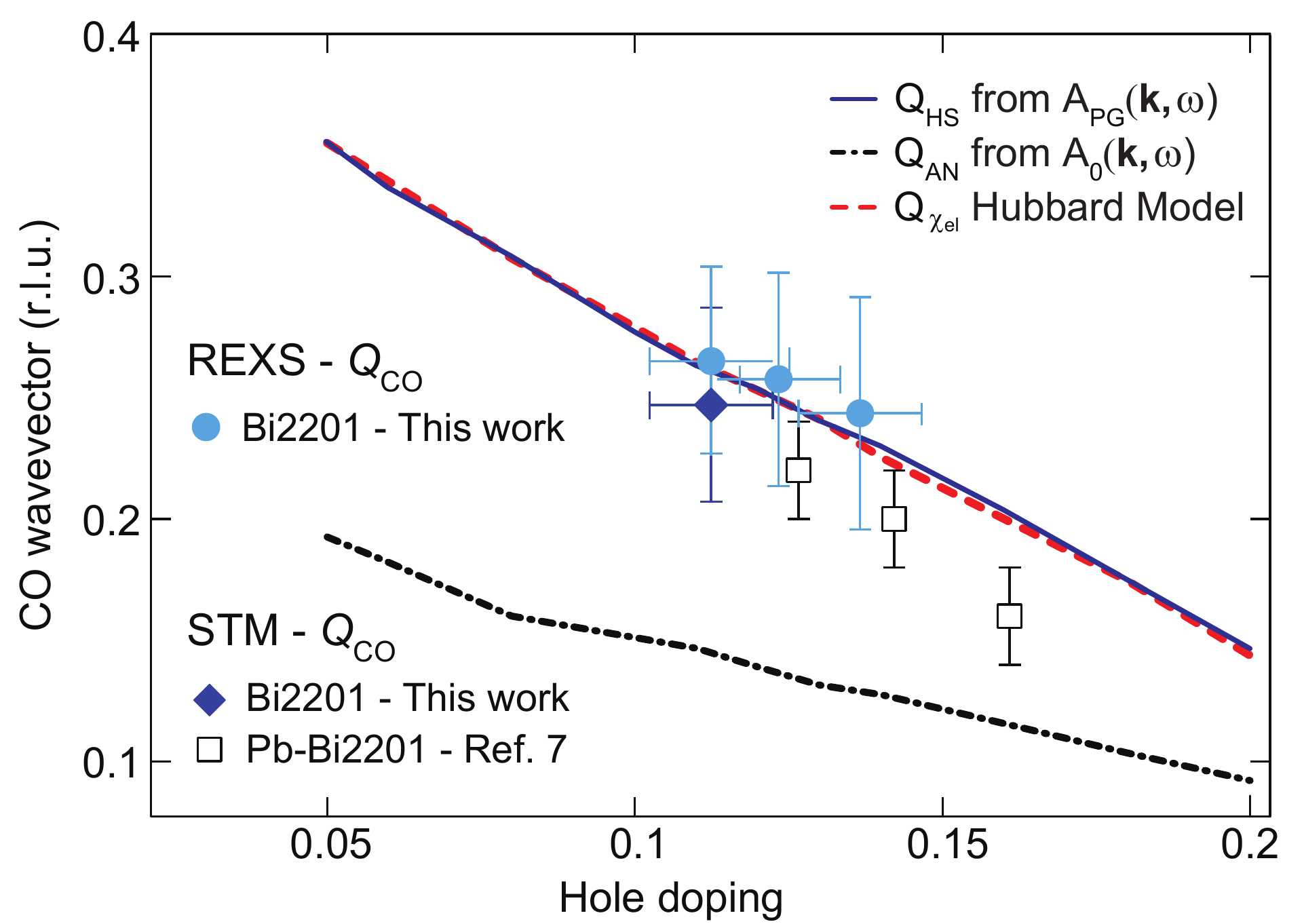,clip=,angle=0,width=1\linewidth}}
\caption{Doping dependence of the charge order wavevector ${Q}_{\mathrm{CO}}$ as determined by REXS and STM on Bi2201 [this work and \cite{wise2008}]; note that bars represent peak widths, and not errors. Also shown are evolution of the Fermi surface-derived wavevectors $Q_{AN}$ (antinodal nesting) and $Q_{HS}$ (arc tips) measured from the ARPES spectral function $A(\mathbf{k}, \omega)$, as well as the doping dependent wavevector ${Q}_{{\chi}_{el}}$ from the Hubbard-model-based electronic susceptibility \cite{Eder}.}
\label{Susc_doping_fig}
\end{figure}

In Fig.\,\ref{Susc_doping_fig} we report the doping dependence of the charge-order wavevector ${Q}_{\mathrm{CO}}$ as seen experimentally, as well as ${Q}_{AN}$ and ${Q}_{HS}$ as obtained from the spectral function ${A}_{0}(\mathbf{k}, \omega)$ and ${A}_{\mathrm{PG}} (\mathbf{k}, \omega)$ for the non-interacting and interacting cases, respectively. The ${T}_{c}$-to-doping conversion for the experimental points is taken from previous studies on La-substituted Bi2201 \cite{Ando2000}; for Pb-substituted Bi2201 \cite{wise2008} this correspondence might be altered because Pb may contribute holes as well. The mechanism based on electron-hole scattering between AN excitations, with wavevector ${Q}_{AN}$, proves to be inadequate throughout the whole doping range. On the other hand, both the wavevector magnitude ${Q}_{HS}$ and doping dependent-slope $ \mathrm{d} {Q}_{HS} / \mathrm{d} p $ agree with the Bi2201 experimental data, thereby establishing a direct connection between charge order and HS scattering. To gain further phenomenological insights into a possible link between the ordering of the electronic density and the available charge dynamics, we evaluate the momentum-dependent electronic response (susceptiblity) near the Fermi surface, or $ {\chi}_{\mathrm{el}} (\mathbf{Q}, \Omega) $ (see Supplementary Note 3 for more details). We approximate $ {\chi}_{\mathrm{el}} (\mathbf{Q}, \Omega) $ as a self-convolution of the single-particle Green's function $\mathcal{G} (\mathbf{k}, \omega)$, in line with a similar approach successfully used in the study of magnetic excitations in cuprates \cite{Inosov_2007,Konik_2012,Dean_2013}. Despite the simplicity of our model, the results for $ \mathrm{Re} \lbrace {\chi}_{\mathrm{el}} \rbrace $ along the direction of the experimental charge ordering confirm that there is an enhancement of particle-hole scattering at a wavevector ${Q}_{{\chi}_{el}}$ closely following ${Q}_{HS}$ (dashed red line in Fig.\,\ref{Susc_doping_fig}). This convergence supports the idea that accounting for the empirical role played by the hot spots is of critical importance for future, more quantitative studies of the electronic instability.

The convergence between the real- and reciprocal-space techniques in our study indicates a well-defined length scale and coherence associated with the electronically-ordered ground state. These findings on Bi2201 suggest that the short-ranged charge correlations in Bi-based cuprates \cite{Hoffman2002,howald2003,vershinin2004,wise2008}, and the longer-ranged modulations seen in Y-based \cite{ghiringhelli2012,chang2012,Blanco2013,Blackburn2013} and La-based compounds \cite{tranquada1995,vZimmermann_1998,abbamonte2005}, are simply different manifestations of a generic charge-ordered state (see Ref.\,\onlinecite{daSilvaNeto2013} for related findings on Bi2212). That the experimental ordering wavevectors can be reproduced through the correlation-induced Fermi arcs in the PG state demonstrates a quantitative link between the single-particle fermiology and the collective response of the electron density in the underdoped cuprates.

\bibliography{Bi2201_CDW_REXS_STM}

\vspace{6mm}

We acknowledge S.A. Kivelson, A.-M. Tremblay, S. Sachdev, E.H. da Silva Neto, and A. Yazdani for discussions. This work was supported by the Max Planck -- UBC Centre for Quantum Materials, the Killam, Alfred P. Sloan, Alexander von Humboldt, and NSERC's Steacie Memorial Fellowships (A.D.), the Canada Research Chairs Program (A.D., G.A.S.), NSERC, CFI, and CIFAR Quantum Materials. J.E.H. acknowledges support from the US National Science Foundation grant DMR-0847433. A. S. was funded by the A*STAR fellowship. M. Y. was funded by an NSERC fellowship. Part of the research described in this paper was performed at the Canadian Light Source, which is funded by the CFI, NSERC, NRC, CIHR, the Government of Saskatchewan, WD Canada, and the University of Saskatchewan.

\section{Supplementary Materials}

\renewcommand{\tablename}{Supplementary Table}
\makeatletter
\renewcommand{\fnum@figure}{\figurename~S\thefigure}
\setcounter{figure}{0}
\makeatother
\renewcommand{\theequation}{S\arabic{equation}} 

\subsection{Materials and Methods}

\noindent {\bf Sample preparation.} This study focused on three underdoped Bi$_2$Sr$_{2-x}$La$_x$CuO$_{6+\delta}$ single crystals ($x\!=\!0.8$, $p\!\simeq\!0.115$, UD15K; $x\!=\!0.6$, $p\!\simeq\!0.13$, UD22K; $x\!=\!0.5$, $p\!\simeq\!0.145$, UD30K). The superconducting $T_{\mathrm{c}}\!=\!15$, 22 and 30\,K, respectively, were determined from magnetic susceptibility measurements. The $T_{\mathrm{c}}$-to-doping correspondence is taken from \cite{Ando2000}.
\\
\noindent {\bf Soft X-ray scattering.} Resonant elastic soft X-ray measurements (REXS) were performed: (i) at BESSY -- beamline UE46-PGM-1, using a XUV-diffractometer; and (ii) at the Canadian Light Source -- beamline REIXS, using a 4-circle diffractometer. In REXS, three different UD15K crystals were studied, hereafter labeled S1, S2 and S3, and one sample each for UD22K and UD30K. The photon energy was tuned to the La-${M}_{5}$ ($ h \nu \!=\! 834.7$\,eV) and Cu-${L}_{3}$ ($ h \nu \!=\! 931.5$\,eV) absorption edges. The probing scheme hinges on control of incoming polarization (we measured both $ \sigma $ and $ \pi $ channels), while no outgoing polarization analysis was performed. Reciprocal-space scans were acquired by rocking the sample angle at fixed detector position, in the temperature range 10-300\,K. In all cases samples were pre-oriented using Laue diffraction and mounted with either {\bf a} or {\bf b} axis in the scattering plane. Both \textit{in}- and \textit{ex-situ} cleaving procedures were used, yielding consistent results.
\\
\noindent {\bf Scanning Tunnelling Microscopy.} The experiments were performed on a home-built cryogenic UHV STM at 9 and 40\,K. STM tips were cut from Pt/Ir wire and cleaned by field emission on polycrystalline Au foil prior to the experiments. Bulk crystals were cleaved \textit{in-situ} in cryogenic UHV and immediately inserted into the STM. Topographic images were measured in constant current mode at a fixed sample bias of -200\,mV, and a tunnelling current of 200-250\,pA. Differential conductance spectra were measured out of tunnelling feedback at a fixed tip-sample separation using a lock-in technique with an excitation voltage of 5 to 20\,mV and a frequency of 1.115\,kHz. Spectroscopic maps were acquired over a 24-48 hour time frame.
\\
\noindent {\bf Electronic susceptibility calculations.} Single-particle Green's functions $\mathcal{G} (\mathbf{k}, \omega)$ have been constructed using Dyson equation $\mathcal{G} (\mathbf{k}, \omega) \!=\! {[\omega-{\epsilon}_{\mathbf{k}}^{\mathrm{bare}}-{\Sigma}_{\mathrm{PG}} (\mathbf{k}, \omega)]}^{-1}$, with tight binding parameters for the bare band dispersion ${\epsilon}_{\mathbf{k}}^{\mathrm{bare}}$ as follows: $t\!=\!0.4$\,eV, $t'/t\!=\!-0.2$, $t''/t\!=\!0.05$, $t'''\!=\!0$. We adopted the analytic form for the self-energy ${\Sigma}_{\mathrm{PG}} (\mathbf{k},\omega)$ provided in \cite{YRZ} [see also Supplementary Note 3 and Fig.\,S\ref{SE_panels}A-C for more details]. The doping-dependent pseudogap magnitude and the parameters for the incoherent continuum have been respectively calibrated to the experimental hump features in the tunnelling DOS, and to the energy extension of the characteristic dispersion of incoherent spectral weight (also known as `waterfall') as seen by ARPES [see Fig.\,S\ref{SE_panels}E and Fig.\,S\ref{ARPES_SF_model}, B1-B4 and E]. We have subsequently calculated the density-density correlator by Wick-decomposing the full particle-hole propagator into a self-correlation of the interacting Green's function (particle-hole bubble). Analytic continuation onto the real-frequency axis was performed, and an FFT-based algorithm was utilized for fast-computation of the correlation functions.

\subsection{Supplementary Text}
\begin{center}\noindent{\bf Supplementary Note 1 $|$ Resonant X-ray Scattering Results.}\end{center}

\begin{figure}[h!]
\centerline{\epsfig{file=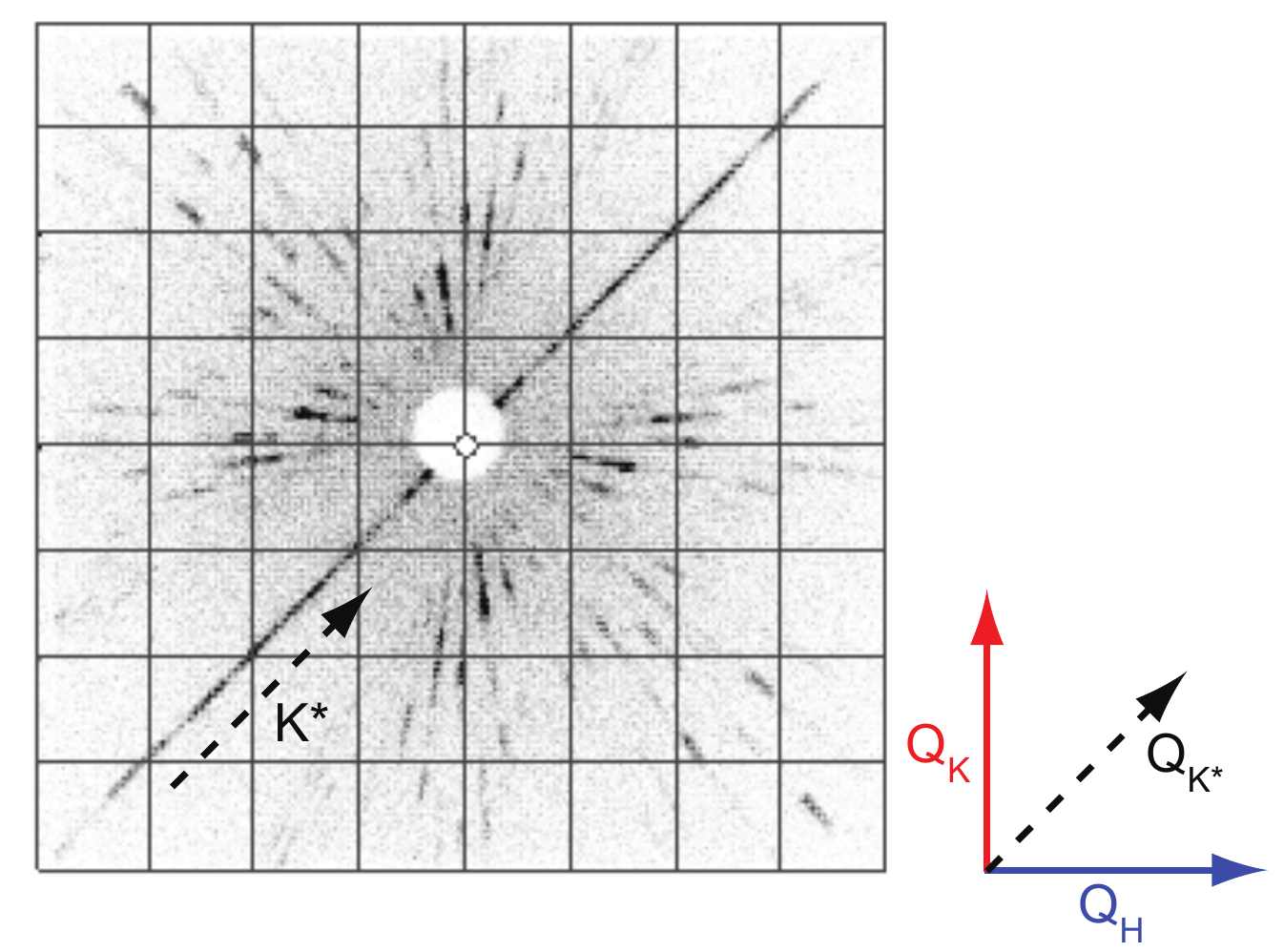,clip=,angle=0,width=0.7\linewidth}}
\vspace{-3mm}
\caption{Main panel: Laue diffraction pattern of sample UD15K-S4; the arrows define our convention for the reciprocal axes \textbf{H} and \textbf{K} we will refer to hereafter.}
\label{Laue}
\end{figure}

\noindent{\bf Definition of reciprocal axes.} The structural symmetry of La-substituted Bi2201 crystals is characterized by an orthorhombic distortion of the tetragonal unit cell, whose new axes are at 45${}^{\circ}$ with respect to the nearest-neighbour Cu-O bond directions (which define the \textbf{a} and \textbf{b} axes) and by the presence of long-range ordered supermodulations along a single axis (\textbf{b*}). From the point of view of symmetry, the \textbf{a} and \textbf{b} axes are equivalent, hence we had to establish an arbitrary convention to distinguish between them in the actual measurement. Supplementary Figure S\ref{Laue} shows a representative Laue pattern of a Bi2201 single-crystal. Besides the lowest-order Bragg reflections, one can see a streak with a high-density of diffracted spots, which corresponds to the direction of the superstructural modulations Q${}_{1}$ (and Q${}_{2}$), commonly associated to the orthorhombic ${\mathbf{Q}}_{{K}^{*}}$ reciprocal axis. The ${\mathbf{Q}}_{H}$ and ${\mathbf{Q}}_{K}$ axes have been defined as indicated by the arrows.

\begin{figure}[h!]
\centerline{\epsfig{file=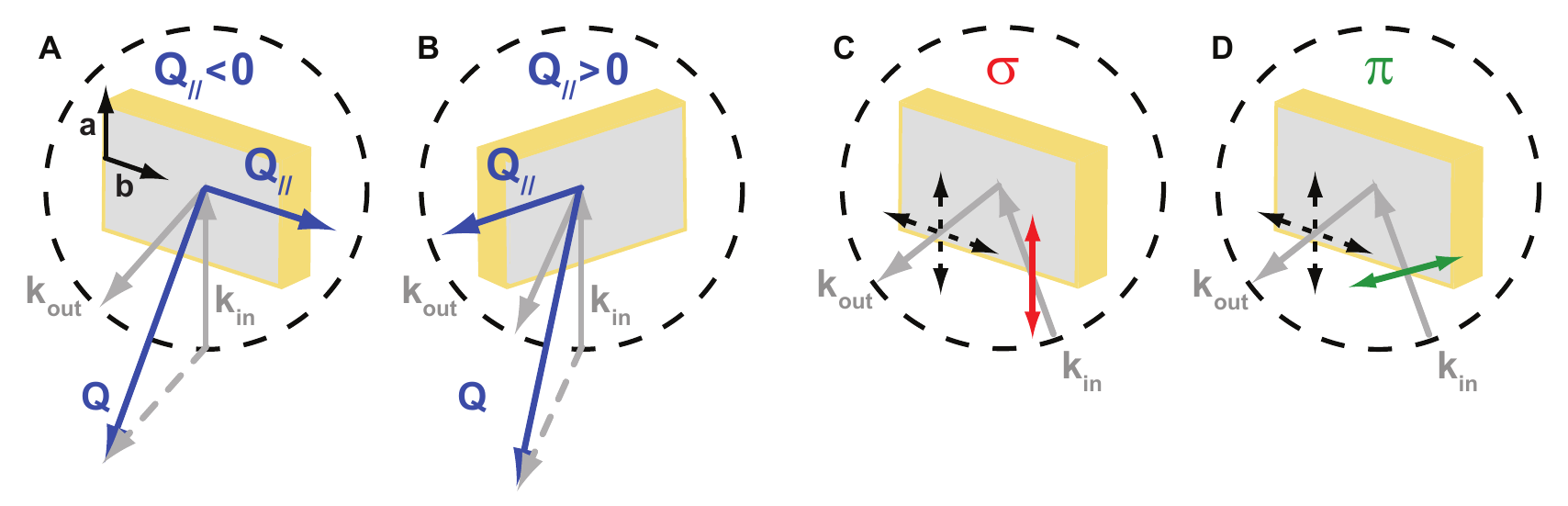,clip=,angle=0,width=0.95\linewidth}}
\caption{Schematics of the experimental geometry in REXS measurements, which illustrate how the negative (\textbf{A}) and positive (\textbf{B}) sign of ${Q}_{\parallel}$ are defined based on the position of the exchanged wavevector \textbf{Q} with respect to the surface normal. (\textbf{C,D}) delineate the two incoming polarization geometries, vertical ($\sigma$) and horizontal ($\pi$), respectively. No polarization analysis on the scattered beam is performed in the current study.}
\label{REXS_geometry}
\end{figure}

\noindent{\bf REXS experimental geometry.} The scattering measurements have been performed in two different geometries, which are conventionally associated to positive and negative wavevectors. This situation is elucidated by Fig.\,S\ref{REXS_geometry}, A and B, for the case of ${Q}_{\parallel}\!<\!0$ and ${Q}_{\parallel}\!>\!0$, respectively. The former case corresponds to ${\theta}_{in}\!>\!{\theta}_{out}$, and vice versa for the latter (angles are measured from the surface normal).
An additional experimental parameter is light polarization. In the soft X-ray regime control over incoming polarization is straightforward, and two geometries can be used - $\sigma$ or $\pi$ - the former referring to having the polarization perpendicular to the scattering plane [see Fig.\,S\ref{REXS_geometry}C], while for the latter the light polarization vector lies in such plane (Fig.\,S\ref{REXS_geometry}D). In the same energy range it is much more difficult and less efficient to select the outgoing polarization, therefore for this study no polarization analysis of the scattered light was performed in any of the measurements.

\begin{figure}[t!]
\centerline{\epsfig{file=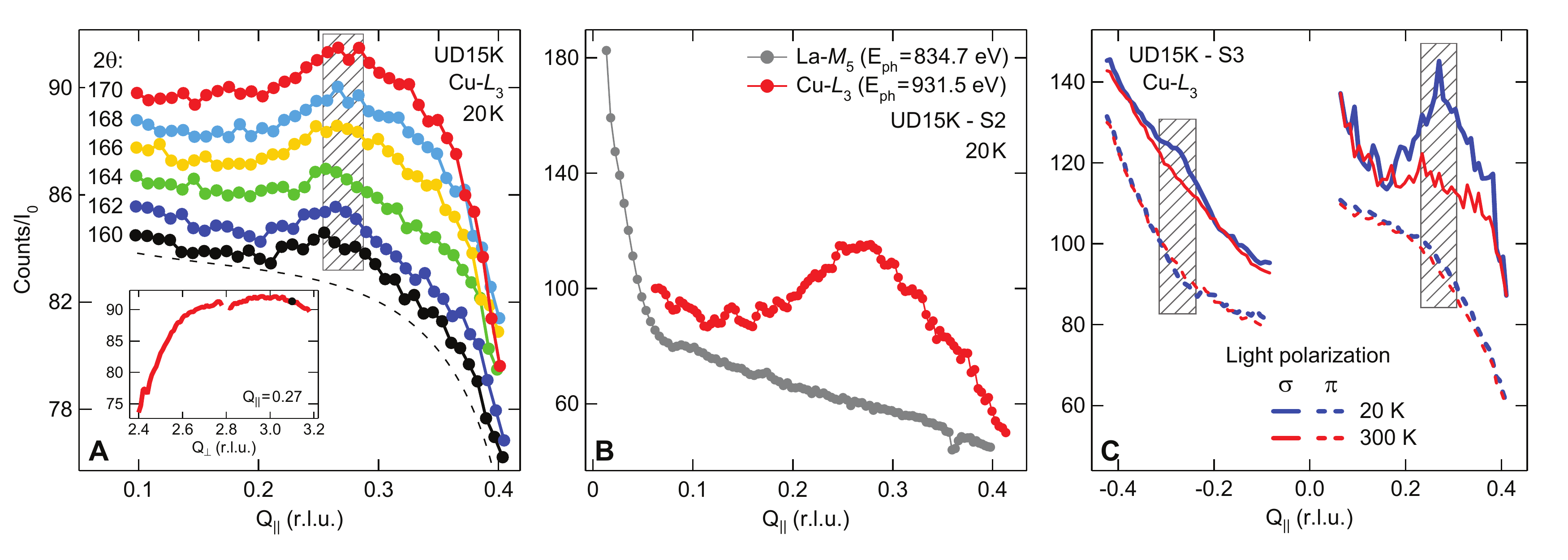,clip=,angle=0,width=1\linewidth}}
\caption{(\textbf{A}) Projection along ${\mathbf{Q}}_{\parallel}$ of REXS $\theta$-scans (rocking curves), taken at the Cu-${L}_{3}$ edge for different values of detector angle 2$\theta$ (profiles are vertically offset for clarity). The inset shows the ${\mathbf{Q}}_{\perp}$ dependence of the scattering signal at ${\mathbf{Q}}_{\parallel}\!=\!0.27$ (no fluorescence subtracted, the black circle denotes the ${\mathbf{Q}}_{\perp}$ value at which all ${\mathbf{Q}}_{\parallel}$-scans were performed). (\textbf{B}) REXS scans at the La-${M}_{5}$ (834.7\,eV) and Cu-${L}_{3}$ (931.5\,eV) edges. (\textbf{C}) Dependence of the scattered intensity on the incoming light polarization ($\sigma$ and $\pi$) at low (20\,K) and high temperature (300\,K), at the Cu-${L}_{3}$ edge. In panels \textbf{A} and \textbf{C}, the gray stripes mark the range where the charge-ordering (CO) peak was observed.}
\label{tth_pol_edge}
\end{figure}

\noindent{\bf Experimental evidence for planar modulations in the CuO${}_{2}$ layers.} Fig.\,S\ref{tth_pol_edge} elucidates the various aspects that motivate the assignment of the detected peaks in REXS to a modulation of the electronic charge within the CuO${}_{2}$ planes. Panel (a) shows a series of rocking curves (scans of the sample angle $\theta$), projected onto the planar component of the wavevector (${Q}_{\parallel}$), for different values of the detector angle $2 \theta$. A broader ${Q}_{\perp}$-dependence (i.e., probing periodicities along the \textbf{c} axis) of the scattering signal near the CO wavevector (${Q}_{\parallel}\!\sim\!0.265$) is plotted in the inset panel. The presence of a peak structure at different detector angles, and the associated weak modulation along ${Q}_{\perp}$ implies the two-dimensional nature of the underlying CO. 
Fig.\,S\ref{tth_pol_edge}B shows two ${Q}_{\parallel}$-scans, taken at the Cu-${L}_{3}$ and La-${M}_{5}$ edges (photon energies were defined based on the maximum of the absorption signal) at 10\,K. The absence of any features at the La-edge reveals that the charge modulation is confined to the CuO${}_{2}$ layers. 
Fig.\,S\ref{tth_pol_edge}C shows the temperature and light polarization dependence of the momentum scans, which indicate that the CO signal is maximum when the incoming light is $\sigma$-polarized. This implies that the charge modulation originates primarily from the intermediate states that can be reached in this geometry, i.e. Cu-${d}_{x^2-y^2}$. Also, throughout the rest of the discussion, unless otherwise specified, it is assumed that all reciprocal space scans have been performed at a fixed detector angle of $ 2 \theta\!=\!{167}^{\circ} $, and at the Cu-${L}_{3}$ edge resonance. 
\begin{figure}[h!]
\centerline{\epsfig{file=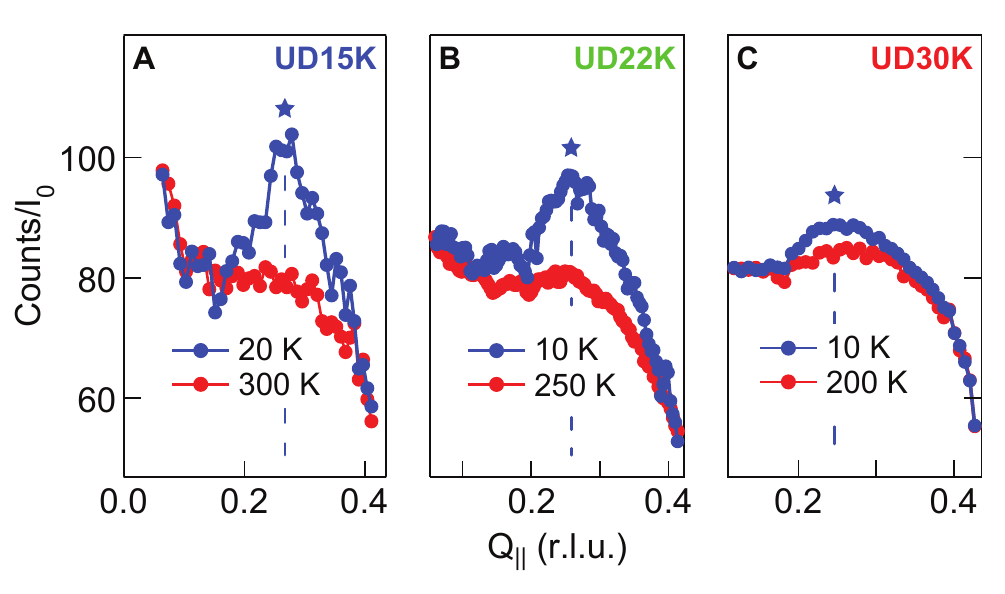,clip=,angle=0,width=1\linewidth}}
\vspace{-2mm}
\caption{Low- and high-temperature scattering scans for the doping levels investigated with REXS: (\textbf{A}) UD15K; (\textbf{B}) UD22K; (\textbf{C}) UD30K. Data have been taken at the Cu-${L}_{3}$ resonance ($h \nu\!=\!931.5$\,eV). Blue stars indicate the location of the charge-ordering wavevector ${Q}_{\mathrm{CO}}$.}
\label{High_low_T_data}
\end{figure}

\begin{figure}[t!]
\centerline{\epsfig{file=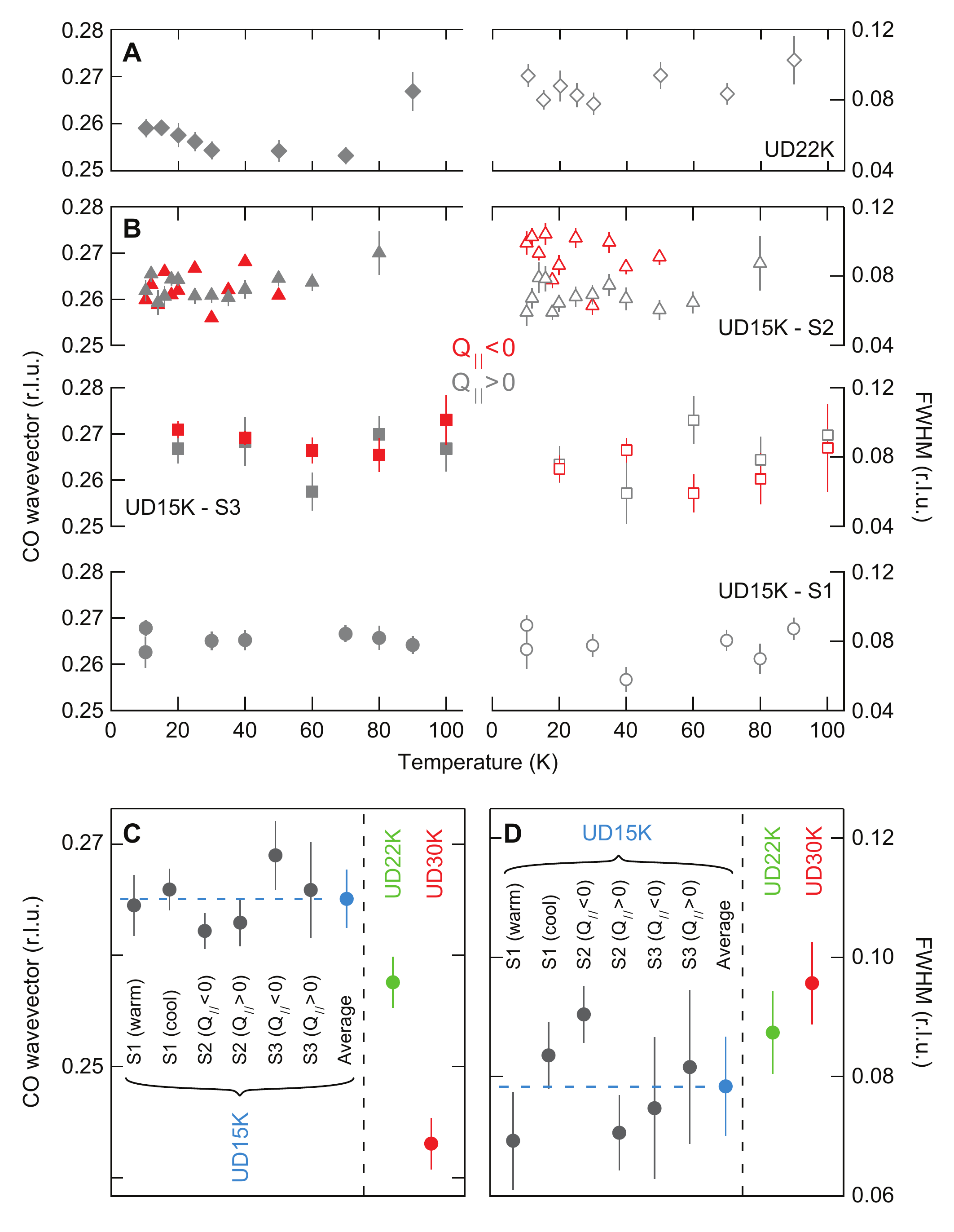,clip=,angle=0,width=1\linewidth}}
\vspace{-3mm}
\caption{(\textbf{A}) Summary of the temperature-dependent results for the CO wavevector and full-width-at-half-maximum (FWHM) for the UD22K sample, in the range $T\!<\!100$\,K. (\textbf{B}) Same as (\textbf{A}), for the various UD15K samples measured. (\textbf{C}) Statistical analysis of the CO wavevector: the gray markers (bars) represent the temperature-averaged values (errors) within each dataset in UD15K, with the blue marker (and dashed line) being the overall average over different samples and geometries; the red and green markers are the single points for UD22K (averaged over temperatures) and UD30K, respectively. (\textbf{D}) Same as (\textbf{C}), for the FWHM.}
\label{Pos_width}
\end{figure}

\noindent{\bf Doping-dependent CO wavevector and correlation lengths.} 
Fig.\,S\ref{High_low_T_data} displays the high- and low-temperature REXS scans for the different doping levels investigated, which are also illustrative of the doping-dependence of the ordering wavevector and peak intensity, both decreasing with increasing hole doping. Figs.\,S\ref{Pos_width}, A and B, show the peak position (left) and FWHM (right) for the various samples and doping levels investigated. We have restricted the temperature axis to the range $T\!<\!100$\,K, where these parameters can be more reliably extracted in virtue of a better signal-to-noise ratio. For the UD15K, note that red (gray) markers are used for ${Q}_{\parallel}\!<\!0$ (${Q}_{\parallel}\!>\!0$).
The bottom panels (Figs.\,S\ref{Pos_width}, C and D) report the results of a statistical analysis applied to the data shown in the top panels, with averaged parameter values and error bars extracted for each dataset (gray markers for UD15K and red markers for UD22K), and further averaged for the UD15K samples to provide a single-valued final estimate for this doping level (blue markers). Overall, we get FWHMs in the range 0.08 to 0.1 (in reciprocal lattice units), which correspond in turn to correlation lengths of 6 to 8 unit cells, or approximately 20-30\,\AA.
\begin{figure}[h!]
\centerline{\epsfig{file=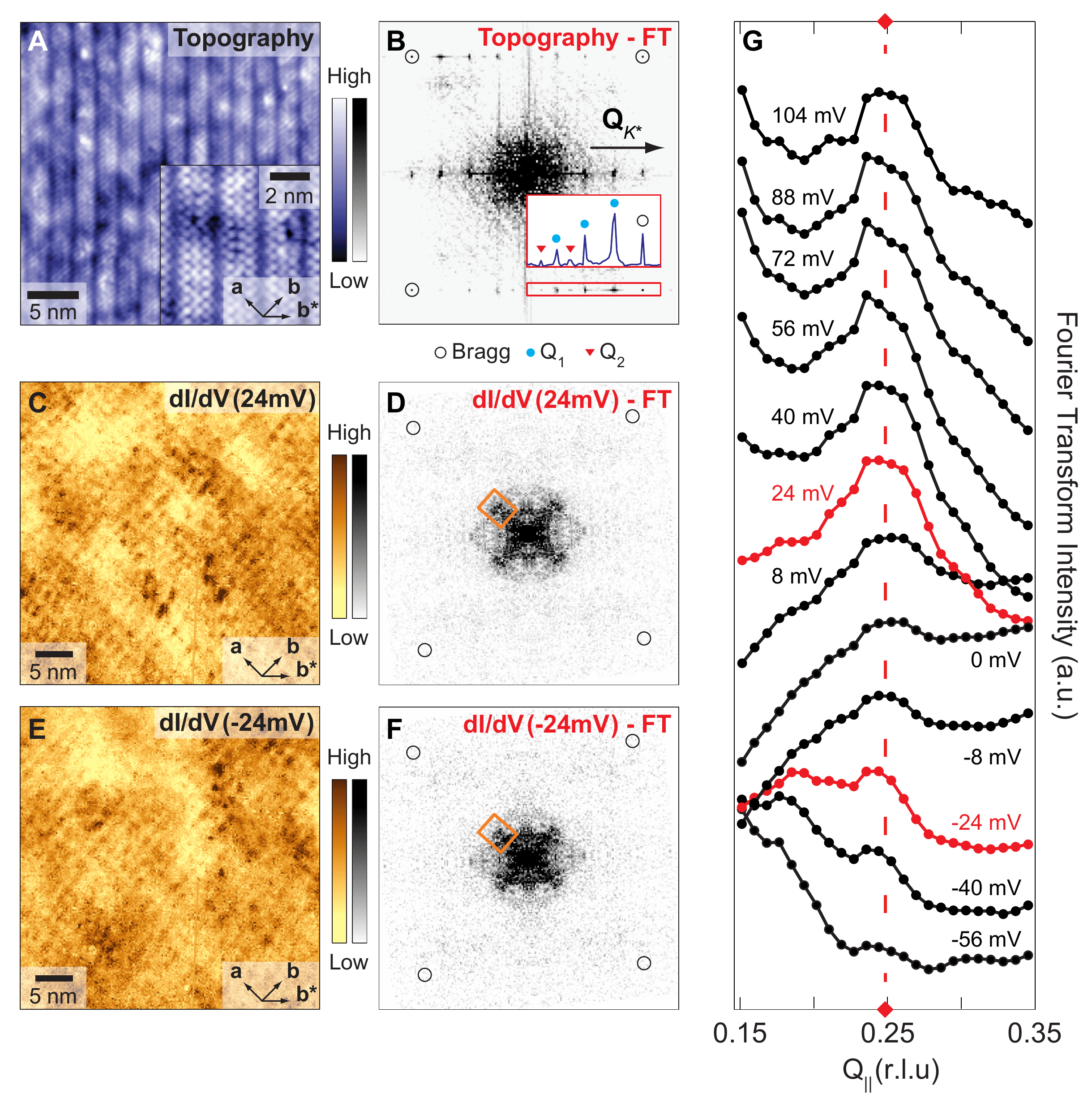,clip=,angle=0,width=1\linewidth}}
\caption{(\textbf{A}) STM topography with atomic resolution over a 29\,nm region ($T\!=$\,4.5\,K, -200\,mV and 20\,pA). The inset shows the supermodulation in higher resolution (8\,nm, -1.2\,V, 100\,pA) (\textbf{B}) Fourier transform (FT) of (\textbf{A}) [hollow circles indicate Bragg vectors $(\pm 1, 0)$ and $(0, \pm 1)$]. The inset shows a linecut through the Bragg, Q${}_{1}$ and Q${}_{2}$ peaks as indicated by black circles, blue dots, and red triangles, respectively. Note that Q${}_{1}$ and Q${}_{2}$ are the structural supermodulation peaks along the orthorhombic reciprocal axis ${\mathbf{Q}}_{{K}^{*}}$ as defined in Ref.\,\onlinecite{Rosen_Comin}. (\textbf{C,E}) $\mathrm{d}I/\mathrm{d}V$ map at respectively 24 and -24\,mV bias over a 29\,nm region (9\,K, -200\,mV and 250\,pA). A checkerboard modulation with period $\sim\!4{a}_{0}$ is clearly seen in real space. (\textbf{D,F}) Symmetrized FT of (\textbf{C,E}), respectively. (\textbf{G}) Stack of linecuts of (\textbf{D}), with vertical offset for clarity, in the range indicated by the orange rectangle in (\textbf{D,F}), as a function of bias voltage $V$ and wavevector ${Q}_{\parallel}$, expressed in reciprocal lattice units [red traces correspond to the bias values used in (\textbf{D,F})]. A charge-order peak can be seen around ${Q}_{\parallel}\!\sim\!0.25$ in the full range of bias voltages spanning from -56 to 104\,mV.}
\label{STM_conductance}
\end{figure}
\vspace{10mm}
\begin{center}\noindent{\bf Supplementary Note 2 $|$ Scanning Tunnelling Microscopy Results.}\end{center}

\noindent
Fig.\,\ref{STM_conductance}A shows the topographic map $T(\mathbf{r})$, acquired at 4.5\,K on a field of view of approximately $29\!\times\!29$\,nm, which images the topmost BiO layer (the natural cleavage plane in these materials). The well-known structural supermodulation Q${}_{1}$ can be already seen in the form of corrugated ripples, but is best visualized in the Fourier-transformed map $\tilde{T}(\mathbf{q})$ as strong satellite spots (see blue dots in Fig.\,\ref{STM_conductance}B) forming a line oriented at 45${}^{\circ}$ with respect to the \textbf{a} and \textbf{b} axes. Notably, the second supermodulation Q${}_{2}$ is also seen in $\tilde{T}(\mathbf{q})$ (red triangles), consistent with our previous findings by ARPES and low-energy electron diffraction \cite{phil,Rosen_Comin}. The conductance map $\mathrm{d}I/\mathrm{d}V(\mathbf{r},V\!=\!24\,\mathrm{mV})$ and its Fourier transform (Fig.\,\ref{STM_conductance}C,D) show a charge-modulated pattern. The latter is shown in Fig.\,\ref{STM_conductance}E to be present for an ample range of bias voltages \textit{V} -- 8 to 104\,mV -- and to be nondispersive with \textit{V}, ruling out the possibility that it arises from quasiparticle scattering, which instead would possess a distinctive dependence on \textit{V} \cite{Hoffman_QPI}.

\begin{center}\noindent{\bf Supplementary Note 3 $|$ Model Green's function and particle hole-propagator.}\end{center}

\begin{figure}[h!]
\centerline{\epsfig{file=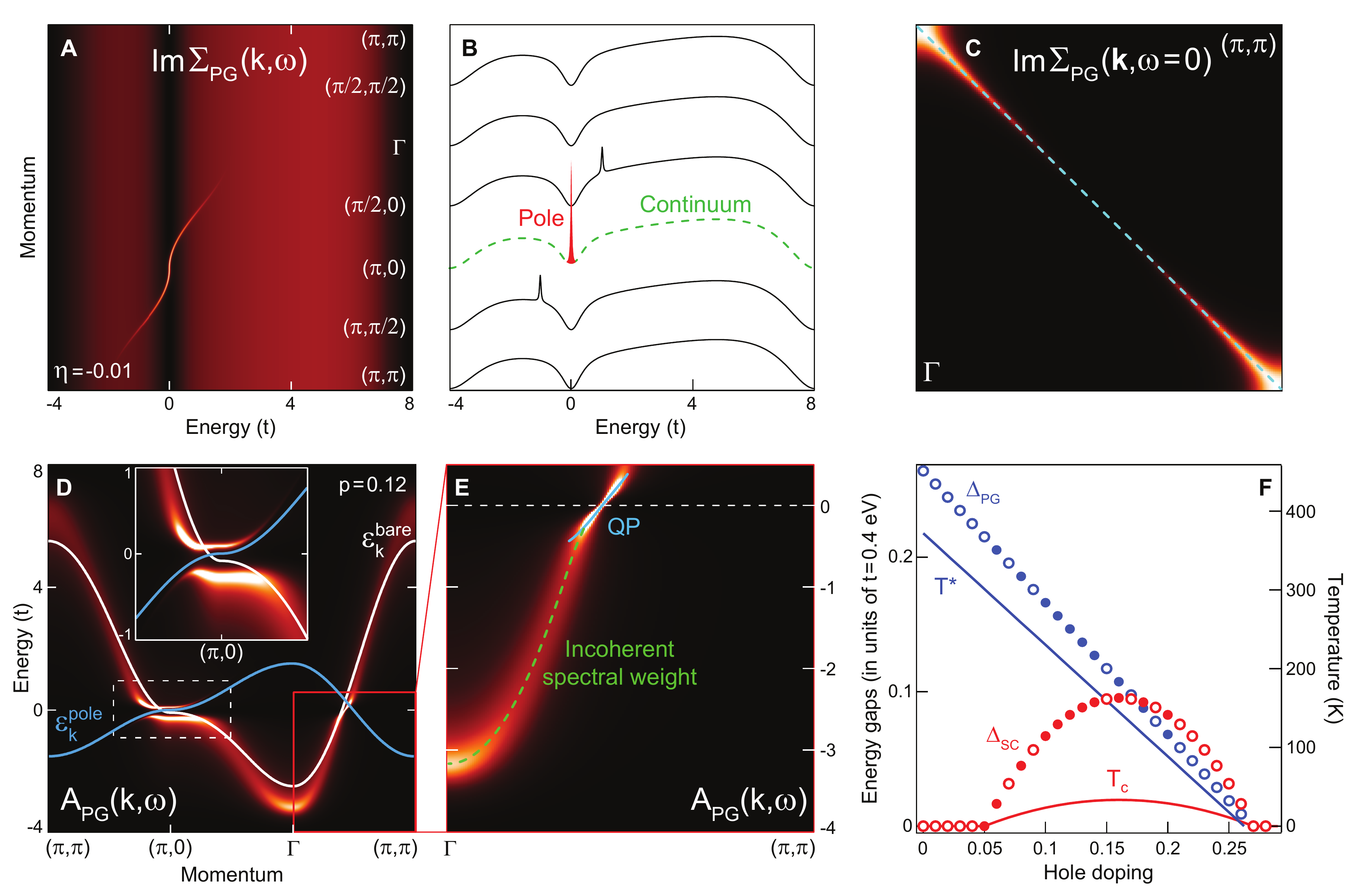,clip=,angle=0,width=1\linewidth}}
\caption{(\textbf{A}) Momentum and energy dependence of the imaginary part of the self-energy $ {\Sigma}_{\mathrm{PG}} (\mathbf{k}, \omega) $, for the full energy range used in this model. (\textbf{B}) Stack of constant-momentum cuts of Im ${\Sigma}_{\mathrm{PG}} (\mathbf{k}, \omega)$ at the high symmetry-points specified in panel (\textbf{A}). (\textbf{C}) Im ${\Sigma}_{\mathrm{PG}} (\mathbf{k}, \omega\!=\!0)$, showing that the line of zeros for Re ${\Sigma}_{\mathrm{PG}} (\mathbf{k}, \omega)$ coincides with the antiferromagnetic zone boundary (dashed light blue line). (\textbf{D}) Momentum and energy depdendence of the spectral function $A (\mathbf{k}, \omega)$, with overlaid the pole in the self-energy (light blue curve) and the bare band (white curve); the inset is an enlarged view near ${E}_{F}$ and the antinode $(\pi,0)$. (\textbf{E}) Nodal cut of $A (\mathbf{k}, \omega)$, which highlights the crossover from coherent excitations (quasiparticles) near the Fermi energy, onto incoherent, broad excitations at higher energies. (\textbf{F}) Diagram of the energy- and temperature-scales used in our model; full symbols mark the doping levels which we considered in the numerical calculations.}
\label{SE_panels}
\end{figure}

\noindent{\bf The underlying phenomenological self-energy.} The basis for our self-energy ${\Sigma}_{\mathrm{PG}} (\mathbf{k}, \omega)$ is derived from the parametrized version provided in \cite{Eder}, which we extended to more doping values than its original formulation (which was specific to the cases $p\!=\!0$, 0.12 and 0.24). Using this self-energy, we can then proceed to evaluate the single-particle propagator, or the retarded Green's function, as: $\mathcal{G} (\mathbf{k}, \omega) \!=\! {(\omega-{\epsilon}_{\mathbf{k}}^{\mathrm{bare}}-{\Sigma}_{\mathrm{PG}} (\mathbf{k}, \omega))}^{-1}$.
This particular form for ${\Sigma}_{\mathrm{PG}} (\mathbf{k}, \omega)$, obtained by extrapolating to all momenta the exact diagonalization results on a single-band $t\!-\!t'\!-\!U$ Hubbard model, incorporates sharp features -- in the form of a dispersing pole (see red peak in Fig.\,S\ref{SE_panels}B) - and broader continua (see dashed green profile in Fig.\,S\ref{SE_panels}B): ${\Sigma}_{\mathrm{PG}} (\mathbf{k}, \omega)\!=\!{\Sigma}_{\mathrm{pole}} (\mathbf{k}, \omega)+{\Sigma}_{\mathrm{cont}} (\mathbf{k}, \omega)$. The dispersing pole is defined as ${\Sigma}_{\mathrm{pole}} (\mathbf{k}, \omega)\!=\! {\vert{\Delta}_{\mathbf{k}}^{\mathrm{PG}}\vert}^{2} {(\omega+{\epsilon}_{\mathbf{k}}^{\mathrm{pole}})}^{-1}$, where $ {\Delta}_{\mathbf{k}}^{\mathrm{PG}} = {\Delta}_{0}^{\mathrm{PG}} (\cos ({k}_{x}) - \cos ({k}_{y})) $ and $ {\epsilon}_{\mathbf{k}}^{\mathrm{pole}}\!=\!2{t}^{\mathrm{pole}}(p)\times (\cos ({k}_{x}) + \cos ({k}_{y})) + {\mu}_{\mathrm{pole}}$ is the reversed nearest-neighbor hopping (plus a constant offset). The dispersion parameter of the self-energy pole ${t}^{\mathrm{pole}}$ is taken from \cite{YRZ}. In general, we retained the functional form of the pole and the continua (with the only addition of a slope component, see Fig.\,S\ref{SE_panels}B), and slightly adjusted the underlying parameters in order to guarantee optimal matching with the experimental spectral functions. Our underlying bare band parameters for the dispersion in the $({k}_{x},{k}_{y})$-plane are as follows: $t\!=\!0.4$\,eV, $t'/t\!=\!-0.2$, $t''/t\!=\!0.05$, $t'''\!=\!0$. The doping-dependent chemical potential $\mu$ has been defined using Luttinger sum rule.
\noindent
The overall momentum-energy map (along high-symmetry directions) is shown in Fig.\,S\ref{SE_panels}A, while a stack of constant-momentum slices at high-symmetry points is plotted in Fig.\,S\ref{SE_panels}B. Note the vanishing of the pole residue $ {\Delta}_{\mathbf{k}}^{\mathrm{PG}} $ along the nodal direction $\Gamma \rightarrow (\pi,\pi)$. In the case where ${\mu}_{\mathrm{pole}}\!=\!0$, which is what we assumed in our model, the additional line of zeros for Re $\mathcal{G} (\mathbf{k}, \omega\!=\!0)$ [which arises from a diverging ${\Sigma}_{\mathrm{PG}} (\mathbf{k}, \omega\!=\!0)$] coincides with the antiferromagnetic (AFM) zone-boundary (see Fig.\,S\ref{SE_panels}C), and is \textit{de facto} responsible for the formation of the Fermi arcs within this framework. The so-called 'hot-spots' are also here defined as the points where the non-interacting Fermi surface intersect the AFM zone-boundary, or equivalently as the loci in momentum space where $ {\epsilon}_{\mathbf{k}}^{\mathrm{bare}}\!=\!{\epsilon}_{\mathbf{k}}^{\mathrm{pole}}\!=\!0$.

\noindent
Such mechanism is highlighted in the inset of Fig.\,S\ref{SE_panels}D, which shows a zoom-in of the spectral function $A (\mathbf{k}, \omega)\!=\! (-1/\pi) \: \mathrm{Im} \mathcal{G} (\mathbf{k}, \omega)$ around $(\pi,0)$. The Green's function and the self-energy are related in such way that the quasiparticle band ${\epsilon}_{\mathbf{k}}^{\mathrm{QP}}$ and the pole band ${\epsilon}_{\mathbf{k}}^{\mathrm{pole}}$ cannot cross each other. This avoided crossing drives the 'back-bending' of ${\epsilon}_{\mathbf{k}}^{\mathrm{QP}}$ as it approaches the crossing between ${\epsilon}_{\mathbf{k}}^{\mathrm{bare}}$ and ${\epsilon}_{\mathbf{k}}^{\mathrm{pole}}$ (see white and blue curves, respectively) and causes the opening of a 'pseudogap' near the antinode. The latter also inherits its unconventional momentum-dependence from the pole residue $ {\Delta}_{\mathbf{k}}^{\mathrm{PG}} $. Note that, when only the sharp dispersing pole is considered, this model coincides with the RVB-derived self-energy originally proposed in \cite{YRZ}, in which case the derived spectral function is purely coherent, and therefore not normalized due to the absence of the incoherent features, which are instead observed experimentally. Such features are better seen in the main panel of Fig.\,S\ref{SE_panels}D, and appear particularly prominent around the $\Gamma$ point, in agreement with experiments. This apparent band of incoherent excitations along the nodal direction (shown in more detail in Fig.\,S\ref{SE_panels}E), often termed the 'waterfall', in our model bottoms down around 1-1.3\,eV, in good agreement with experimental data on different compounds \cite{Meevasana2007,FournierNP}.

\begin{figure}[t!]
\centerline{\epsfig{file=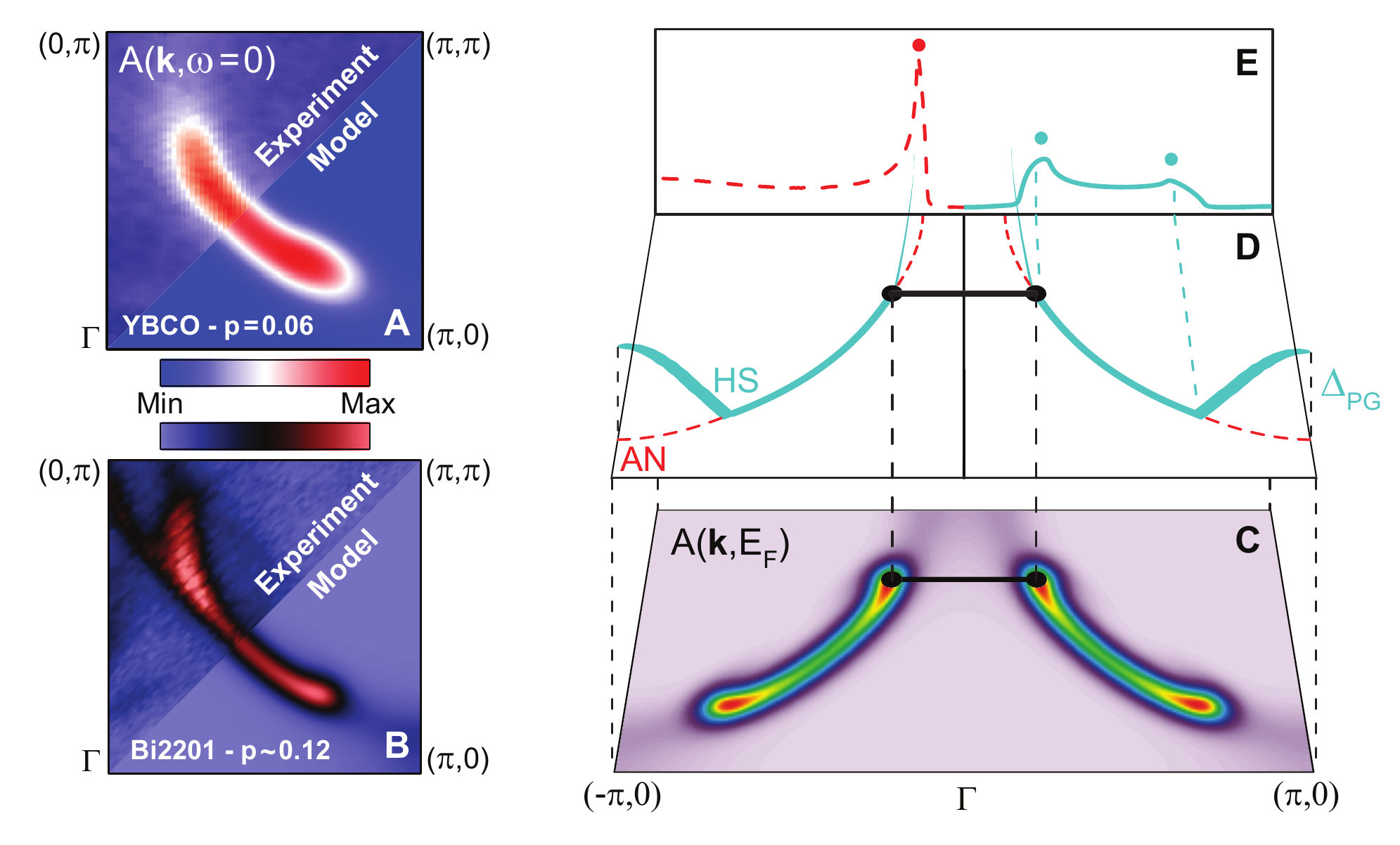,clip=,angle=0,width=1\linewidth}}
\caption{(\textbf{A}) Comparison between experimental (YBCO, $p\!=\!0.06$, from \cite{FournierNP}) and calculated Fermi surface; a Gaussian convolution has been applied to the calculated map to mimic momentum-resolution effects. (\textbf{B}) Same as (\textbf{A}), for the case of UD La-substituted Bi2201 ($p\!\sim\!0.12$, from \cite{phil}). (\textbf{C}) calculated FS for $p\!=\!0.12$, with black circles marking the location of the hot-spots (HS). (\textbf{D}) Energetics of lowest-energy excitations near the FS in presence of the pseudogap; the blue curve is the \textbf{k}-dependent PG, the red dashed profile is the non-interacting FS. (\textbf{E}) Cumulative densities of states $\tilde{\rho} ({k}_{x})$ for the non-interacting (red dashed) and interacting case (full blue); the ${k}_{x}$-positions of enhanced cumulative DOS are marked with circles.}
\label{cumDOS}
\end{figure}

\begin{figure}[b!]
\centerline{\epsfig{file=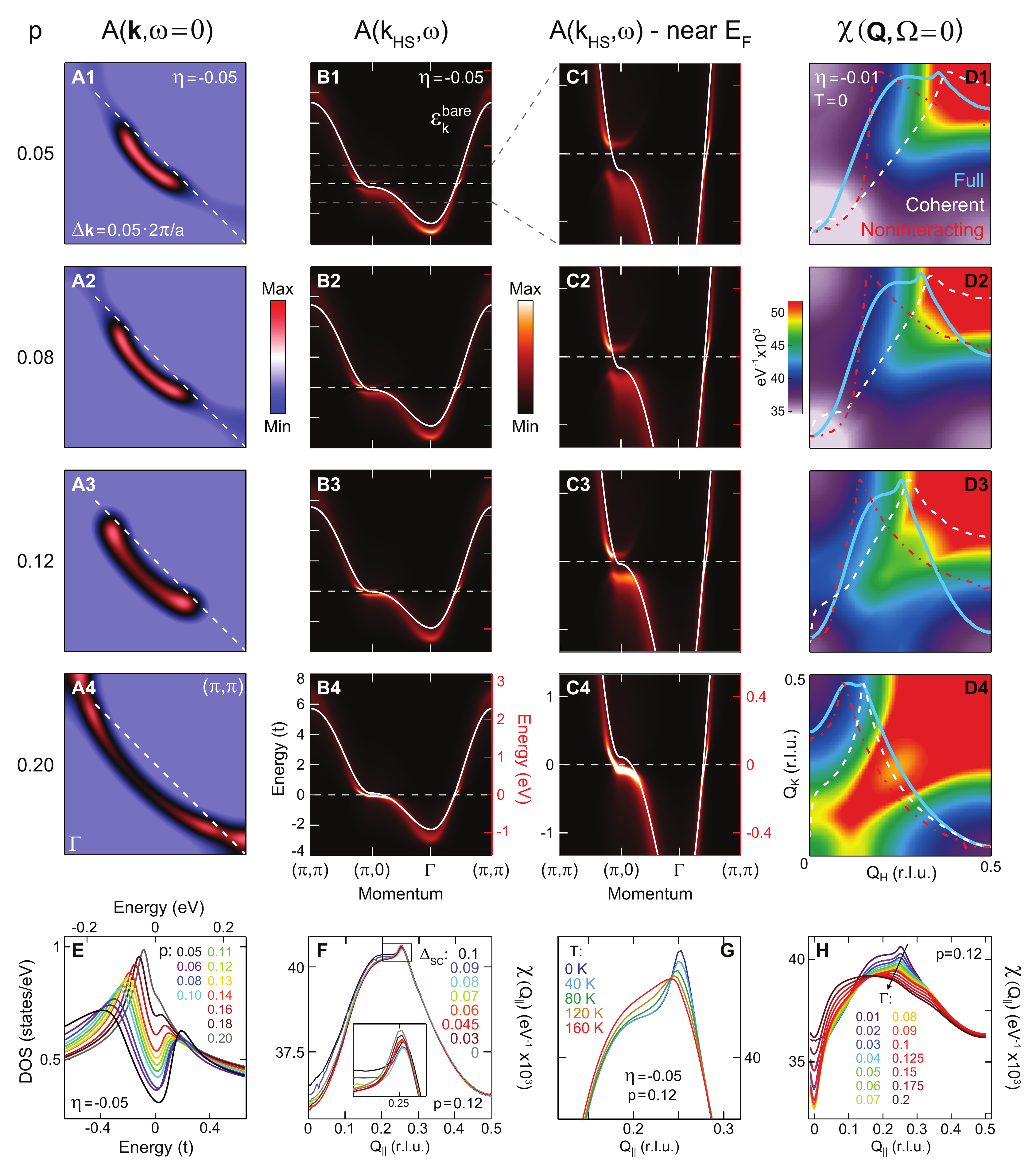,clip=,angle=0,width=1\linewidth}}
\caption{(\textbf{A1-A4}) Fermi surfaces for different hole-doping levels ($p\!=\!0.05$, 0.08, 0.12 and 0.2), obtained as a constant-energy slice of the spectra function at the Fermi energy $A (\mathbf{k}, \omega\!=\!0)$, followed by convolution to an isotropic momentum-resolution Gaussian function with $\Delta {k}_{x}\!=\!\Delta {k}_{y}\!=\!0.05 \: \pi / a$. (\textbf{B1-B4}) Momentum-energy maps of $A (\mathbf{k}, \omega)$ along high-symmetry directions within the first Brillouin-zone (the white curves represent the bare band dispersion). (\textbf{C1-C4}) Same as (\textbf{B1-B4}), but zoomed in around ${E}_{F}$. (\textbf{D1-D4}) Resulting zero-frequency ($\Omega\!=\!0$) susceptibility profiles, calculated as the particle-hole bubble of the full Green's function $\mathcal{G} (\mathbf{k}, \omega)$, at $T\!=\!0$. Overlaid are ${Q}_{\parallel}$ cuts of $ {\chi}_{\mathrm{el}} (\mathbf{Q}, \Omega\!=\!0)$ for different underlying Green's functions: (i) full (light blue full line); (ii) coherent part only (dashed white); (iii) noninteracting (dotted-dashed red). (\textbf{E}) Doping-dependent density of states. (\textbf{F,G,H}) Dependence of $ {\chi}_{\mathrm{el}}$ on superconducting gap $ {\Delta}_{SC} $ (\textbf{F}), temperature $T$ (\textbf{G}), and scattering rate $\Gamma$ (\textbf{H}), for $p\!=\!$0.12.}
\label{ARPES_SF_model}
\end{figure}

\noindent
Fig.\,S\ref{SE_panels}F shows the energy- and temperature-scales we used in the subsequent analysis. Most of the results only use the doping dependence of ${\Delta}_{0}^{\mathrm{PG}}$, which we calibrated by comparing the pseudogap-induced humps in the density of states to the expected trend for the pseudogap magnitude \cite{hufner}.
\\
\noindent{\bf Fermi surface transformation and cumulative density of states.}
The flexibility of the model we just introduced allows achieving an excellent agreement with the experimentally determined spectral functions, as is shown in Fig.\,S\ref{cumDOS}, A and B, for the comparison between a calculated Fermi surface (FS) and the corresponding experimental map in underdoped YBCO \cite{FournierNP} and La-substituted Bi2201 \cite{phil}, respectively. Within the framework of Fermi-arcs, the hot-spots are defined as those points in the first Brillouin zone where zero-energy excitations become gapped out; they are marked by the black circles in Fig.\,S\ref{cumDOS}C, with the underlying intensity map representing a calculated FS for $p\!=\!0.12$. Fig.\,S\ref{cumDOS}D schematizes the mechanism: zero-energy electronic excitations lying on the non-interacting FS (dashed red line) are pushed away from the Fermi level between the antinode (AN) and the HS (the full blue line is the pseudogap profile ${\Delta}_{\mathbf{k}}^{\mathrm{PG}}$).
\noindent
As a consequence, the band topology is altered due to the PG-induced band-bending (see also Fig.\,S\ref{SE_panels}D) and the pile-up of the local (in momentum-space) density of states (DOS) migrates from the AN to the HS. This behavior can be quantitatively captured by computing the cumulative DOS $\tilde{\rho} ({k}_{x})$, defined as follows: 
\begin{equation}
\tilde{\rho} ({k}_{x})\!=\! \int_{- {\Delta}_{\mathrm{PG}}}^{{\Delta}_{\mathrm{PG}}} \mathrm{d} \omega \int_{-2\pi /a}^{2\pi /a} \mathrm{d} {k}_{y} A ({k}_{x}, {k}_{y}, \omega)
\end{equation}
\noindent
which is plotted in Fig.\,S\ref{cumDOS}E for the case with (full blue line) and without (dashed red line) the PG. It is clear that when the latter is turned on, a single peak (at the AN) in $\tilde{\rho} ({k}_{x})$ evolves into a profile with two smaller peaks near the HS, thus setting the ground for enhanced electron-hole scattering between these momentum points. 
\\
\noindent{\bf Doping-dependence of the spectral function.} With this phenomenological self-energy in hand, we then proceed to calculate the full Green's function as a function of doping. The corresponding Fermi surfaces and momentum-energy maps (for $p\!=\!0.05$, 0.08, 0.12 and 0.2) are shown in Fig.\,S\ref{ARPES_SF_model}, A1-A4 and B1-B4. The panels C1-C4 are the same as B1-B4, but zoomed in energy around the Fermi energy. The associated doping-dependent densities of states (DOS) are obtained by tracing out the spectral function over momenta: $ \rho (\omega) = \int \mathrm{d} \mathbf{k} \: A (\mathbf{k}, \omega) $. The resulting profiles are plotted in Fig.\,S\ref{ARPES_SF_model}E, showing the progressive opening of the pseudogap as hole-doping is decreased, in the form of two side-humps around ${E}_{F}$.
\\
\noindent{\bf Calculation of the electronic response.} We made use of our phenomenological Green's function to calculate the low-energy electronic response in Fourier space, as encoded in the particle-hole propagator $ {\mathcal{G}}_{2} (\mathbf{Q}, i \Omega_n) $ ($ \Omega_n $ denotes a bosonic frequency). This quantity, which is generally referred to as the electronic susceptibility ${\chi}_{\mathrm{el}}$, and reduces to the Lindhard function for a non-interacting system, is defined as the retarded charge-charge correlation function:

\begin{equation}
{\chi}_{\mathrm{el}} (\mathbf{Q}, \tau) \propto \int_{0}^{\beta} \mathrm{d \tau} {e}^{i {\Omega}_{n} \tau} \langle {T}_{\tau} \: ({\rho}_{el} (\mathbf{Q},\tau) {\rho}_{el} (- \mathbf{Q},0)) \rangle \end{equation}
\begin{equation}
\hspace{10mm} = \sum_{\mathbf{k}, \mathbf{k'}, \sigma, \sigma '} \langle {T}_{\tau} \: ({c}_{\mathbf{k},\sigma}^{\dagger} (\tau ) \: {c}_{\mathbf{k+Q},\sigma} (\tau ) \: {c}_{\mathbf{k},\sigma '}^{\dagger} {c}_{\mathbf{k-Q},\sigma '}) \rangle \label{eq:chi_definition}
\end{equation}
\noindent
where ${T}_{\tau}$ is the imaginary-time ordering operator, and the Fourier-transformed electronic density operator $ {\rho}_{el} (\mathbf{Q},\tau) $ has been expanded in terms of charge creation and annihilation operators $ c $ and $ {c}^{\dagger} $.
We now adopt a generalization of a Lindhard function (independent particle) approach to the interacting problem, which consists of rewriting Eq.\,\ref{eq:chi_definition} as a self-convolution of the interacting Green's function $\mathcal{G} (\mathbf{k}, \omega)$. This approach neglects the correlation between the particular electron-$(\mathbf{k}, \sigma)$ and hole-$(\mathbf{k} + \mathbf{Q}, \sigma)$ state, but retains the effects of correlations with all other states, which enter through the single-particle self-energy ${\Sigma}_{\mathrm{PG}} (\mathbf{k}, \omega)$ [in analogy to similar approaches in optical spectroscopy \cite{timusk,Sharapov2005}]. This procedure has been previously used for the study of magnetic excitations in cuprates \cite{Inosov_2007,Konik_2012,Dean_2013}. In more general contexts, the real part $ \mathrm{Re} \lbrace {\chi}_{\mathrm{el}} \rbrace $ is in some cases indicative of the electronic contribution to instabilities of the ionic lattice or of the charge density \cite{Johannes_2008}. In reality, a complete assessment of the system's tendency toward ordering phenomena requires the detailed evaluation of all the coupling terms between the different interconnected degrees of freedom, \textit{in primis} electron-phonon coupling \cite{Qin2010,Bonnoit2012,LeTacon2013} and exchange interactions \cite{Konik_2012,Sachdev2013}.

Our approach is therefore based on expressing the susceptibility as follows:
\begin{equation}
{\chi}_{\mathrm{el}} (\mathbf{Q}, \tau) \propto \sum_{\mathbf{k}, \sigma} \mathcal{G} (\mathbf{k} + \mathbf{Q}, \tau, \sigma) \mathcal{G} (\mathbf{k}, \tau, \sigma) \label{eq:Wick_decomposition} \end{equation}
\noindent
with $\mathcal{G} (\mathbf{k}, \tau, \sigma) = \langle {T}_{\tau} \: ({c}_{\mathbf{k},\sigma} (\tau) \: {c}_{\mathbf{k},\sigma}^{\dagger}) \rangle$.

\noindent
This is now equivalent to approximating the electronic susceptiblity with the particle-hole bubble diagram of the full Green's function $\mathcal{G}$. By Fourier-transforming Eq.\,\ref{eq:Wick_decomposition} into the frequency-domain, we ultimately arrive to a form of $ {\chi}_{\mathrm{el}} (\mathbf{Q}, i {\Omega}_{n}) $ which is conveniently expressed as a frequency-momentum autocorrelation of $ \mathcal{G} $:
\begin{equation}
{\chi}_{\mathrm{el}} (\mathbf{Q}, i {\Omega}_{n}) \propto \frac{1}{\beta} \: \sum_{\mathbf{k}, i {\omega}_{m}, \sigma} \mathcal{G} (\mathbf{k} + \mathbf{Q}, i {\omega}_{m} + i {\Omega}_{n}, \sigma) \: \mathcal{G} (\mathbf{k}, i {\omega}_{m}, \sigma)
\label{eq:chi_definition_freq}
\end{equation}
\noindent
where we have here defined $ \beta\!=\!{(k T)}^{-1} $, with $ {\Omega}_{n}\!=\! 2n \pi / \beta $ and $ {\omega}_{m}\!=\! (2m+1) \pi / \beta $ representing the bosonic and fermionic Matsubara frequencies, respectively. The summation in Eq.\,\ref{eq:chi_definition_freq} is then converted to an integration on the real-frequency axis (analytic continuation) using the substitution: $i {\omega}_{m} \rightarrow \omega + i \eta$ (hereafter $\eta\!=\!0.01$, unless otherwise specified). 
The correlation in the momentum coordinates was performed using Fast-Fourier Transform (FFT); unless otherwise specified, grids of $256 \times 256$ k- and Q-points have been used. The (${k}_{x}$,${k}_{y}$)-momentum ranges were set between $ - \pi / a $ and $ \pi / a $, and periodical boundary conditions (PBCs) have been imposed in the correlation to avoid boundary effects. Values of $\eta\!=\!-0.01$ and $\Delta \omega\!=\!0.005$ (both in units of $t\!=\!0.4$\,eV) have been used to guarantee a proper sampling of the energy axis, which is crucial especially near ${E}_{F}$ where excitations become more coherent, and the related features in $ \mathcal{G} (\mathbf{k},\omega) $ get sharper. 
The validity and applicability of the numerical code has been benchmark-tested on the two-dimensional electron gas (2DEG), whose analytic solution is known.
\begin{figure}[t!]
\centerline{\epsfig{file=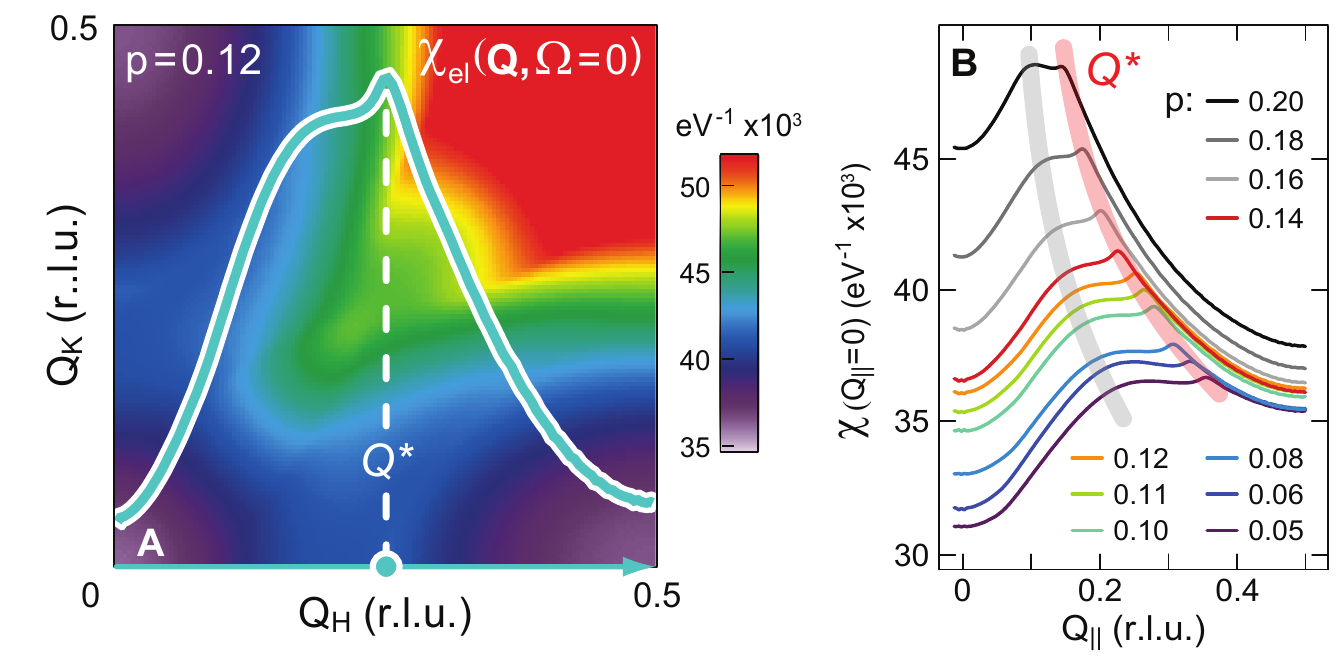,clip=,angle=0,width=1\linewidth}}
\caption{(\textbf{A}) Zero-frequency, zero-temperature electronic susceptibility for $p\!=\!0.12$; overlaid is the cut along ${\mathbf{Q}}_{H}$ (full blue line), and the local maximum at ${Q}_{{\chi}_{el}}\!\sim\!0.26$ highlighted. (\textbf{B}) Stack of doping-dependent ${\chi}_{\mathrm{el}} ({Q}_{\parallel})$ profiles, with red and gray guides-to-the-eye tracking the contributions from scattering between HS and AN, respectively.}
\label{Susc_fig_sup}
\end{figure}
\noindent
The zero-temperature, zero-frequency electronic susceptibility $ {\chi}_{\mathrm{el}}^{T=0} (\mathbf{Q}, \Omega\!=\!0) $ for $p\!=\!0.12$ is plotted in Fig.\,S\ref{Susc_fig_sup}A, with overlaid the cut along ${\mathbf{Q}}_{H}$ (light blue trace). The shape of this profile is determined by two contributions: (i) a peak at ${Q}_{{\chi}_{el}}$ corresponding to enhanced particle-hole scattering between the hot-spots; and (ii) particle-hole excitations across the PG from the AN regions, which populate the ``hump'' at lower $Q$ values. This profile shows a local maximum at ${Q}_{{\chi}_{el}}\!=\!0.252$, which closely matches the experimental $Q_{\mathrm{CO}}$ observed in Bi2201 (UD15K) and the ${Q}_{HS}$ vector introduced before. In addition, two-dimensional maps of the electronic susceptibility have been calculated for a series of different doping levels, and the corresponding cuts along ${\mathbf{Q}}_{\parallel}$ are plotted in Fig.\,S\ref{Susc_fig_sup}B. The doping evolution of the HS and AN components is indicated by, respectively, the red and gray guides overlaid on the plot.

\noindent{\bf Doping-dependence of the electronic response.} Figs.\,S\ref{ARPES_SF_model}, D1-D4, display the two-dimensional maps of the zero-frequency, zero-temperature electronic susceptibility as a function of momentum, for different doping levels. Overlaid on top of the color maps are the profiles along ${Q}_{H}$ calculated in different configurations: (i) $ \chi ({Q}_{\parallel}, \Omega\!=\!0) $ from the full Green's function (full light blue profile); (ii) $ {\chi}_{\mathrm{coh}} ({Q}_{\parallel}, \Omega\!=\!0) $ from the coherent part of Green's function only (dashed white); (iii) $ {\chi}_{\mathrm{nonint}} ({Q}_{\parallel}, \Omega\!=\!0) $ from the noninteracting Green's function (dash-dotted red). For illustrative convenience, the various profiles are normalized to their minimum and maximum; also note that a proper normalization of $ {\chi}_{\mathrm{coh}} ({Q}_{\parallel}, \Omega\!=\!0) $ is not possible, due to the missing incoherent contributions.
These susceptibility profiles reflect important aspects of the underlying fermiology. The peaks in $ {\chi}_{\mathrm{coh}} ({Q}_{\parallel}, \Omega\!=\!0) $ are mainly controlled by scattering between the hot-spots, while the non-interacting susceptibility has a quasi-1D divergence at lower momenta, which is driven by nesting at the antinodes. As it can be seen at all doping levels, the full susceptibility bears both tendencies, with the antinodal scattering being progressively suppressed as hole-doping is reduced, and correspondingly the pseudogap increased. The presence of some remnant contribution from the antinodes is not surprising, as all electronic states contribute, and not only the zero-energy single-particle excitations at the Fermi surface. Nonetheless, the contibution from lower-lying states decreases as $1/E$ and is therefore less prominent, in agreement with the observed doping dependence of the (pseudo)gapped antinodal component in $ \chi ({Q}_{\parallel}, \Omega\!=\!0) $. Also, the incoherent spectral weight in $\mathcal{G}(\mathbf{k},\omega)$ contributes to the slowly-varying (with ${Q}_{\parallel}$) background in $\chi$, in contrast with the sharper profiles of ${\chi}_{\mathrm{coh}}$ and ${\chi}_{\mathrm{nonint}}$.

\noindent{\bf Dependence of the electronic response on superconducting gap, temperature and scattering rate.} With this machinery in hand, we can calculate the dependence on various other control parameters. In principle it is possible to extend our Green's function to the case where a superconducting ground state is present. In the simplest approximation, this requires adding the following term to the self-energy: ${\Sigma}_{\mathrm{SC}} (\mathbf{k}, \omega) = {\vert{\Delta}_{\mathbf{k}}^{\mathrm{SC}}\vert}^{2} {(\omega + {\epsilon}_{\mathbf{k}}^{\mathrm{bare}} + {\Sigma}_{\mathrm{PG}})}^{-1} $, where ${\Delta}_{\mathbf{k}}^{\mathrm{SC}}$ is the d-wave superconducting order parameter, and ${\epsilon}_{\mathbf{k}}^{\mathrm{bare}}$ is the full bare band dispersion. The total self-energy then reads $ \Sigma\!=\!{\Sigma}_{\mathrm{PG}} + {\Sigma}_{\mathrm{SC}} $. At the level of susceptibility, this introduces an additional diagram besides the particle-hole bubble, which arises from the self-correlation of the anomalous propagator $ F (\mathbf{k}, \omega) $ (see \cite{YRZ} and \cite{Abanov1999} for more details on the subject). The resulting profiles for $p\!=\!0.12$, as a function of the superconducting gap ${\Delta}_{0}^{\mathrm{SC}}$, are shown in Fig.\,S\ref{ARPES_SF_model}F. As it can be seen, near the wavevector ${Q}_{{\chi}_{el}}$ that maximizes $ \chi ({Q}_{\parallel}, \Omega\!=\!0) $, there is hardly any dependence on ${\Delta}_{0}^{\mathrm{SC}}$ -- a maximum reduction of $ \chi ({Q}_{{\chi}_{el}}, \Omega\!=\!0) $ of 0.5\% between the profiles for ${\Delta}_{0}^{\mathrm{SC}}\!=\!0$ and 0.1 is seen to occur.

\noindent
Temperature-dependent suceptibility profiles are shown in Fig.\,S\ref{ARPES_SF_model}G. As temperature is increased, the peak at ${Q}_{{\chi}_{el}}$ is progressively reduced and broadened, whereas the antinodal component at lower \textit{Q} rises due to thermally-enhanced access to excitations around the antinode. The weak temperature dependence of ${Q}_{{\chi}_{el}}$ vs. temperature is a direct consequence of our model, which requires the underlying self-energy ${\Sigma}_{\mathrm{PG}}$ -- and therefore the spectral function -- to be temperature-independent. This assumption implies, in particular, that the poles of the spectral function, and hence also the hot-spots, will not change location as a function of temperature. This is not inconsistent with the empirical observation of temperature-dependent Fermi-arc length, which was recently explained to arise from the variation in temperature of a single parameter (the scattering rate $\Gamma$), without needing the rest of the bandstructure parameters ($t$, $t'$, $t''$, ${t}_{\mathrm{pole}}$, and ${\Delta}_{\mathrm{PG}}$) to change with temperature \cite{Reber_2012}.

\noindent
Lastly, the dependence of $ \chi ({Q}_{\parallel}, \Omega\!=\!0) $ on the single-particle inverse lifetime, or scattering rate $\Gamma$ (which is introduced through the imaginary part of the complex frequency using the substitution $\omega \rightarrow \omega+i\Gamma$), shown in Fig.\,S\ref{ARPES_SF_model}H reveals that the $\Gamma$-induced broadening of the Green's function drives a similar broadening in the main features of the susceptibility, in a fashion which resembles the one observed with temperature. This might suggest that susceptibility-driven electronic instabilities are potentially suppressed by incipient disorder and/or impurity scattering. 

\end{document}